\makeatletter
\def\ps@ta{%
   \renewcommand{\@oddhead}{\begin{minipage}{13cm}
        \footnotesize\itshape  A slightly different version of this manuscript will appear
 in Mathematical Physics, Analysis and Geometry. \end{minipage}
                                }%
   \renewcommand{\@evenhead}{}%
   \renewcommand{\@oddfoot}{\hfil\footnotesize\textsf{\bfseries\thepage}\hfil}%
   \renewcommand{\@evenfoot}{}%
}%
\makeatother
\documentclass[11pt]{amsart}
\usepackage{a4}
\usepackage{amssymb}
\usepackage{enumerate}
\usepackage{latexsym}
\usepackage{psfrag}
\usepackage[dvips]{graphicx}

\newcommand{\ZZ}{{\mathbb Z}}
\newcommand{\PP}{{\mathbb P}}
\newcommand{\RR}{{\mathbb R}}
\newcommand{\CC}{{\mathbb C}}
\newcommand{\NN}{{\mathbb N}}

\newcommand{\EE}{{\mathbb E}}

\newtheorem{theorem}{Theorem}[section]
\newtheorem{lemma}[theorem]{Lemma}
\newtheorem{prop}[theorem]{Proposition}
\newtheorem{coro}[theorem]{Corollary}
\theoremstyle{definition}
\newtheorem{definition}[theorem]{Definition}
\theoremstyle{remark}
\newtheorem{remark}[theorem]{Remark}
\newtheorem*{exmp*}{Example}

\newcommand{\supp}{{\mathop{\mathrm{supp}}}}
\newcommand{\nr}[1]{\vert #1\vert}

\DeclareMathOperator{\Ric}{\mathrm{Ric}}

\newcommand{\calB}{\mathcal{B}}
\newcommand{\calD}{\mathcal{D}}
\newcommand{\calE}{\mathcal{E}}
\newcommand{\calF}{\mathcal{F}}
\newcommand{\calG}{\mathcal{G}}
\newcommand{\calH}{\mathcal{H}}
\newcommand{\calI}{\mathcal{I}}
\newcommand{\calJ}{\mathcal{J}}
\newcommand{\calK}{\mathcal{K}}
\newcommand{\calM}{\mathcal{M}}
\newcommand{\calN}{\mathcal{N}}
\newcommand{\calV}{\mathcal{V}}
\newcommand{\calX}{\mathcal{X}}

\renewcommand\qedsymbol{$\Box$}

\newcommand{\tr}{{\mathrm{tr}}}

\newcommand{\Hm}[1]{\leavevmode{\marginpar{\tiny%
$\hbox to 0mm{\hspace*{-0.5mm}$\leftarrow$\hss}%
\vcenter{\vrule depth 0.1mm height 0.1mm width \the\marginparwidth}%
\hbox to
0mm{\hss$\rightarrow$\hspace*{-0.5mm}}$\\\relax\raggedright #1}}}

\newcounter{smalllist}

\begin{document}
\title[Groupoids, von Neumann Algebras and the IDS]
{Groupoids, von Neumann Algebras and the Integrated Density of States}
\author[D.~Lenz]{Daniel Lenz}
\author[N.~Peyerimhoff]{Norbert Peyerimhoff}
\author[I.~Veseli\'c]{Ivan Veseli\'c}
\address[D.~Lenz \& I.~Veseli\'c ]{Fakult\"at f\"ur Mathematik, D-09107 \ TU Chemnitz, Germany }
\urladdr{www.tu-chemnitz.de/mathematik/mathematische\_physik/}
\urladdr{www.tu-chemnitz.de/mathematik/schroedinger/}
\address[N.~Peyerimhoff]{Dept.~of Mathematical Sciences, Durham University, UK}
\urladdr{http://www.maths.dur.ac.uk/~dma0np/}
\date{\today, \jobname.tex. 
}


\maketitle 
\thispagestyle{ta}

\begin{abstract}
We study spectral properties of random operators in the general
setting of groupoids and von Neumann algebras. In particular, we
establish an explicit formula for the canonical trace of the von
Neumann algebra of random operators and define an abstract density of
states.

While the treatment applies to a general framework we lay special
emphasis on three particular examples: random Schr\"odinger operators
on manifolds, quantum percolation and quasi crystal Hamiltonians. For
these examples we show that the distribution function of the abstract
density of states coincides with the integrated density of states
defined via an exhaustion procedure.

\end{abstract}

\date{\today, \time, \jobname.tex}

\section{Introduction}
The aim of this paper is to review and present a unified treatment of
basic features of random (Schr\"odinger) operators using techniques
from Connes' noncommutative integration theory and von Neumann
algebras \cite{Connes-1979}. Particular emphasis will be laid on an
application of the general setting to the example of

\begin{itemize}

\item a group action on a manifold proposed by two of the authors
  \cite{PeyerimhoffV-2001}.

\end{itemize}

This example merges and extends two situations, viz periodic
operators on manifolds as studied first by Adachi/Sunada
\cite{AdachiS-1993} and random Schr\"odinger operators on $\RR^d$ or
$\ZZ^d$ as studied by various people (s. below) starting with the work
of Pastur \cite{Pastur-1971}.  In the first situation a key role is
played by the geometry of the underlying manifold. In the second
situation, the crucial ingredient is the randomness of the
corresponding potential. 

\smallskip

We also apply our discussion to two more examples:

\begin{itemize}
  
\item random operators on tilings and Delone sets whose mathematically
  rigorous study goes back to Hof \cite{Hof-1993} and Kellendonk
  \cite{Kellendonk-1995}).

\item random operators on site-percolation graphs, see
  e.g.~\cite{deGennesLM-59b,ChayesCFST-86,BiskupK-01a,Veselic-05b}.

\end{itemize}

As for the above three examples let us already point out the following
differences: in the first example the underlying geometric space is
continuous and the group acting on it is discrete; in the second
example the underlying geometric space is discrete and the group
acting on it is continuous; finally, in the third example both the
underlying geometric space and the group acting on it are discrete.

The use of von Neumann algebras in the treatment of special random
operators is not new. It goes back at least to the seminal work of
\v Subin on almost periodic operators \cite{Shubin-1979}.  These points
will be discussed in more detail next.

\smallskip

Random Schr\"odinger operators arise in the quantum mechanical
treatment of disordered solids. This includes, in particular, periodic
operators, almost periodic operators and Anderson type operators on
$\ZZ^d$ or $\RR^d$ (cf.~the textbooks \cite{CyconFSK-1987,
Kirsch-89a,CarmonaL-1990, PasturF-1992, Stollmann-2001}).  In all
these cases one is given a family $(H_\omega)$ of selfadjoint
operators $H_\omega$ acting on a Hilbert space $\calH_\omega$, indexed
by $\omega$ in a measure space $(\Omega, \mu)$ and satisfying an
equivariance condition with respect to a certain set of unitary
operators $(U_i)_{i\in I}$.

While specific examples of these cases exhibit very special spectral
features, there are certain characteristics shared by all
models. These properties are as follows. (In parentheses we give a
reference where the corresponding property is established.)

\begin{itemize}
\item [(P1)] Almost sure constancy of the spectral properties of
 $H_\omega$ given some ergodicity condition. In particular, the
 spectrum $\Sigma$ is nonrandom (Theorem \ref{constant}).
\item[(P2)] Absence of discrete spectrum (Corollary \ref{discrete})
and, in fact, a dichotomy (between zero and infinity) for the values
of the dimensions of spectral projections.
\item[(P3)] A naturally arising von Neumann algebra (Section 3) with a
canonical trace $\tau$, to which the random operators are affiliated
(Section 4).
\item[(P4)] A measure  $\rho$, called the {\em density of states},
 governing global features of the family $(H_\omega)$, in particular,
 having $\Sigma$ as its support (Proposition \ref{dandos}). 
 This measure is related to the trace of the von Neumann algebra. 
\end{itemize}

Let us  furthermore single out the following point, which we show for the three abovementioned examples:
\begin{itemize}
\item[(P5)] A local procedure to calculate $\rho$ via an exhaustion
  given some amenability condition. This is known as Pastur-\v Subin trace
  formula.  It implies the self-averaging property of the density of
  states (discussed for the examples mentioned above in Sections 
  \ref{PVresults}, \ref{Quasicrystal}, \ref{Percolation}).
\end{itemize}

Let us now discuss these facts for earlier studied models.  The
interest in property (P5) arouse from the physics of disordered
media. First mathematically rigorous results on the (integrated)
density of states are due to Pastur
\cite{Pastur-1971,Pastur-1972,Pastur-1974}, Fukushima, Nakao and Nagai
\cite{Fukushima-1974,FukushimaNN-1975,FukushimaN-1976,Nakao-1977,
Fukushima-1981}, Kotani \cite{Kotani-1976}, and Kirsch and Martinelli
\cite{KirschM-1982a,KirschM-1982b}. In these papers two different
methods for constructing the integrated density of states (IDS) can be found
(property (P5)). Either one uses the Laplace transform to conclude the
convergence of certain normalized eigenvalue counting functions, or
one analyzes the counting functions directly via the so called
Dirichlet-Neumann bracketing.  In our setting the Laplace transform
method seems to be of better use, since the pointwise superadditive
ergodic theorem \cite{AkcogluK-1981} used in the Dirichlet-Neumann
bracketing approach \cite{KirschM-1982a} has no counterpart in the
(nonabelian) generality we are aiming at.

For the more recent development in the study of the IDS of alloy type
and related models, as well as the results on its regularity and
asymptotic behaviour, see
\cite{CarmonaL-1990,PasturF-1992,Stollmann-2001,Veselic-04a} and the references
cited there.

For almost periodic differential operators on $\RR^d$ and the
associated von Neumann algebras, a thorough study of the above
features (and many more) has been carried out in the seminal papers by
Coburn, Moyer and Singer \cite{CoburnMS-1973} and {\v{S}}ubin
\cite{Shubin-1979}.  Almost periodic Schr\"odinger operators on
$\ZZ^d$ and $\RR^d$ were then studied by Avron and Simon
\cite{AvronS-1982a, AvronS-1983}.

An abstract $C^\ast$-algebraic framework for the treatment of almost
periodic operators was then proposed and studied by Bellissard
\cite{Bellissard-1986,Bellissard-1992}
and Bellissard, Lima and Testard
\cite{BellissardLT-1986}. While these works focus on K-theory and the
so called gap-labeling, they also show (P1)-(P5) for almost periodic
Schr\"odinger type operators on $\RR^d$ and $\ZZ^d$. Let us emphasize
that large parts of this $C^\ast$-algebraic treatment are not confined
to almost periodic operators. In fact, (P1)-(P4) are established there
for crossed products arising from arbitrary actions of locally compact
abelian groups on locally compact spaces $X$.

After the work of Aubry/Andr\'{e} \cite{AubryA-1980} and the short
announcement of Bellissard/Testard in \cite{BellissardT-1982},
investigations in this framework, centered around so called spectral
duality, were carried out by Kaminker and Xia \cite{KaminkerX-1987}
and Chojnacki \cite{Chojnacki-1992}.  A special one-dimensional
version of spectral duality based on \cite{GordonJLS-1997} can also be
found in \cite{Lenz-1999}.

An operator algebraic framework of crossed-products (involving von
Neumann crossed products) can also be used in the study of general
random operators if one considers $\RR^d$ actions together with
operators on $L^2(\RR^d)$ (cf.~\cite{Lenz-1999}).  However, certain of
these models rather use actions of $\ZZ^d$ together with operators on
$L^2(\RR^d)$, like the thoroughly studied alloy or continuous Anderson
type models. This presents a difficulty which was overcome in a work
by Kirsch \cite{Kirsch-1985} introducing a so called suspension
construction, see also \cite{BellissardLT-1986} for related material.
This allows to ``amplify`` these $\ZZ^d$ actions to $\RR^d$ actions
and thus reduce the treatment of (P1)--(P4) in the $\ZZ^d$ case to the
$\RR^d$ case.
\smallskip

In recent years three more classes of examples 
have been considered. These are random operators on manifolds
\cite{Sznitman-1989,Sznitman-1990,PeyerimhoffV-2001,LenzPV-03a,LenzPV-03b},
discrete random operators on tilings
\cite{Hof-1993, Hof-1995,
Kellendonk-1995,Kellendonk-1997,BellissardHZ-2000, LenzS-2001, LenzS-2002},
and random Hamiltonians generated by percolation processes 
\cite{BiskupK-01a,Veselic-05a,Veselic-05b,KirschM-04}.  In
these cases the algebraic framework developed earlier could not be
used to establish (P1)--(P5). Note, however, that partial results
concerning, e.g., (P1) or restricted versions of (P5) are still
available. Note also that continuous operators associated with tilings
as discussed in \cite{BellissardHZ-2000,BenedettiBG-2001} fall within
the $C^\ast$-algebraic framework of
\cite{Bellissard-1986,Bellissard-1992}.
A more detailed analysis of the point spectrum of discrete operators 
associated to tilings and percolation graphs will be carried out in 
\cite{LenzV}.

The model considered in \cite{PeyerimhoffV-2001} includes periodic
operators on manifolds.  In fact, it was motivated by work of Adachi
and Sunada \cite{AdachiS-1993}, who establish an exhaustion
construction for the IDS as well as a representation as a
$\Gamma$-trace in the periodic case. For further investigations related
to the IDS of periodic operators in both discrete and continuous 
geometric settings, see e.g.~\cite{DodziukM-1997,DodziukM-1998,
MathaiY-2002,MathaiSY-2003,DodziukLMSY-03,LenzV}. \smallskip

More precisely, our first example  concerned with {\bf R}andom {\bf S}chr\"odinger
operators on {\bf M}anifolds (RSM)  can be described as follows,see \cite{PeyerimhoffV-2001, LenzPV-03a}:

\begin{exmp*}{\bf (RSM)} {\rm  Let $(X,g_0)$ be the Riemannian  covering 
of a compact Riemannian manifold $M = X / \Gamma$. We assume that
there exists a family $( g_\omega )_{\omega \in \Omega}$ of
Riemannian metrics on $X$ which are parameterized by the elements of a
probability space $(\Omega,\calB_\Omega,\PP)$ and which are uniformly
bounded by $g_0$, i.e., there exists a constant $A \ge 1$ such that 
\begin{equation*}
\frac{1}{A} g_0(v,v) \le g_\omega(v,v) \le A g_0(v,v) \ \ \text{for
all $v \in TX$ and $\omega \in \Omega$}. 
\end{equation*}
 Let $\lambda^\omega$
denote the Riemannian volume form corresponding to the metric
$g_\omega$. We assume that $\Gamma$ acts ergodically on $\Omega$
by measure preserving transformations. The metrics are compatible in
the sense that  for all  $\gamma \in \Gamma$ the corresponding deck transformations 
\begin{equation*}
 \gamma: (X,g_\omega) \to (X,g_{\gamma \omega}) 
\end{equation*}
 are isometries. Then the induced maps
$U_{(\omega,\gamma)}: L^2(X,\lambda^{\gamma^{-1}\omega}) \to
L^2(X,\lambda^\omega)$, $(U_{(\omega,\gamma)} f)(x) = f(\gamma^{-1}x)$
are unitary operators. Based on this geometric setting, we consider a
family $( H_\omega : \omega\in \Omega)$, $H_\omega = \Delta_\omega + V_\omega$,
of Schr\"odinger operators satisfying the following {\em equivariance}
condition
\begin{equation} \label{compcomp} 
H_\omega = U_{(\omega,\gamma)} H_{\gamma^{-1} \omega} U_{(\omega,\gamma)}^*, 
\end{equation} 
for all $\gamma \in \Gamma$ and $\omega \in \Omega$. We also assume
some kind of weak measurability in $\omega$, namely, we will assume
that 
\begin{equation}\label{wm}
\omega \mapsto \langle f(\omega,\cdot), F(H_\omega)
f(\omega,\cdot)\rangle_\omega := \int_X \bar f(\omega,x) \, [F(H_\omega)
f](\omega,x) \, d\lambda^\omega(x)
\end{equation}
is measurable for every measurable $f$ on $\Omega\times X$ with
$f(\omega,\cdot)\in L^2 (X,\lambda^\omega)$, $\omega \in \Omega$, and
every function $F$ on $\RR$ which is uniformly bounded on the spectra
of the $H_\omega$. Note that $L^2 (X, \lambda^\omega)$ considered as a
set of functions (disregarding the scalar product) is independent of
$\omega$. The expectation with respect to the measure $\PP$ will be denoted by $\EE$.}
\end{exmp*}

This example covers the following two particular cases:
\begin{enumerate}[(I)]
\item \label{rsm1} A family of Schr\"odinger operators $( \Delta +
V_\omega)_{\omega \in \Omega}$ on a fixed Riemannian manifold
$(X,g_0)$ with random potentials, see \cite{PeyerimhoffV-2001}.  In
this case the equivariance condition (\ref{compcomp}) transforms into
the following property of the potentials
\begin{equation*} V_{\gamma \omega}(x) = V_\omega(\gamma^{-1}x). \end{equation*}
\item  \label{rsm2}
  A family of Laplacians $\Delta_\omega$ on a manifold $X$ with
random metrics $( g_\omega )_{\omega \in \Omega}$ satisfying some additional assumptions \cite{LenzPV-03a}. 
\end{enumerate}

By the properties of $X$ and $M$ in (RSM), the group $\Gamma$ is
discrete, finitely generated and acts cocompactly, freely and properly
discontinuously on $X$.

\smallskip

In the physical literature the equivariance condition \eqref{compcomp}
is denoted either as equivariance condition, see
e.g.~\cite{Bellissard-1986}, or as ergodicity of operators
\cite{Pastur-80,Kirsch-89a}, where it is assumed that the
measure preserving transformations are ergodic.  From the
probabilistic point of view this property is simply the stationarity
of an operator valued stochastic process.

\medskip

It is our aim here to present a groupoid based approach to (P1)--(P4)
covering all examples studied so far. This includes, in particular, the case of random operators on manifolds, the tiling case and the percolation case.  

Our
framework applies also to Schr\"odinger operators on hyperbolic space
(e.g. the Poissonian model considered in \cite{Sznitman-1989}).
However, our proof of (P5) does not apply to this setting because of the
lack of amenability of the isometry group.

\medskip

For the example (RSM) with amenable group action $\Gamma$, we will prove (P5)
in Section \ref{PVresults}. Note that case (I) of (RSM) includes the models
treated earlier by the suspension construction. Thus, as a by product
of our approach, we get an algebraic treatment of (P5) for these
models.
As mentioned already, our results can also be applied to further
examples.  Application to tilings is discussed in
\cite{LenzS-2001,LenzS-2002}. There, a uniform ergodic type theorem
for tilings along with a strong version of Pastur-\v{S}ubin-formula
(P5) is given.  The results also apply to random operators on
percolation graphs.  In particular they provide complementary
information to the results in
\cite{Veselic-05a,Veselic-05b,KirschM-04}, where the
integrated density of states was defined rigorously for site and edge
percolation Hamiltonians.  
A basic discussion of these examples and
the connection to our study here is given in Sections
\ref{Quasicrystal} and \ref{Percolation} respectively. This will in
particular show that (P5) remains valid for these examples.  
For further details we refer to the cited literature.
The results also apply to random operators on foliations (see
\cite{Kordyukov-95} for related results). 

\medskip

Our approach is based on groupoids and Connes theory of noncommutative
integration \cite{Connes-1979}. Thus, let us conclude this section
by sketching the main aspects of the groupoid framework used in this
article. The work \cite{Connes-1979} on noncommutative integration
theory consists of three parts. In the first part an abstract version
of integration on quotients is presented. This is then used to
introduce certain von Neumann algebras (viz.~von Neumann algebras of
random operators) and to classify their semifinite normal
weights. Finally, Connes studies an index type formula for
foliations. We will be only concerned with the first two parts of
\cite{Connes-1979}.

The starting point of the noncommutative integration is the fact that
certain quotients spaces (e.g., those coming from ergodic actions) do
not admit a nontrivial measure theory, i.e., there do not exist many
invariant measurable functions. To overcome this difficulty the
inaccessible quotient is replaced by a nicer object, a
groupoid. Groupoids admit many transverse functions, replacing the
invariant functions on the quotient. In fact, the notion of invariant
function can be further generalized yielding the notion of random
variable in the sense of \cite{Connes-1979}.  Such a random variable
consists of a suitable bundle together with a family of measures
admitting an equivariant action of the groupoid.  This situation gives
rise to the so called von Neumann algebra of random operators and it
turns out that the random operators of the form $(H_\omega)$
introduced above are naturally affiliated to this von Neumann
algebra. Moreover, each family $(H_\omega)$ of random operators gives
naturally rise to many random variables in the sense of Connes.
Integration of these random variables in the sense of Connes yields
quite general proofs for main features of random operators.  In
particular, an abstract version of the integrated density of states is
induced by the trace on the von Neumann algebra.
\medskip

{\small  \textbf{ Acknowledgments:} It is a pleasure to thank W.~Kirsch and
P.~Stollmann for various stimulating discussions on random
operators and hospitality at Ruhr-Universit\"at Bochum, respectively TU Chemnitz. 
This work was supported in part by the DFG through the SFB 237, the Schwerpunktprogramm
1033, and grants Ve 253/1 \& Ve 253/2 within the Emmy-Noether Programme.}

\section{Abstract setting for basic geometric objects:\\
groupoids and random variables}

In this section we discuss an abstract generalization for the
geometric situation given in example (RSM). The motivation for this
generalization is that it covers many different settings at once, such
as tilings, percolation, foliations, equivalence relations and our
concrete situation, a group acting on a metric space.

Let us first introduce some basic general notations which are
frequently used in this paper. For a given measurable space
$(S,\calB)$ we denote the set of measures by $\calM(S)$ and the
corresponding set of measurable functions by $\calF(S)$. The symbol
$\calF^+(S)$ stands for the subset of nonnegative measurable
functions. $M_f$ denotes the operator of multiplication with a
function $f$.

We begin our abstract setting with a generalization of the action
of the group $\Gamma$ on the measurable space
$(\Omega,\calB_\Omega)$. This generalization, given by $\calG = \Omega
\times \Gamma$ in the case at hand,  is called a groupoid.  The main
reason to consider it
 is the fact that it serves as a
useful substitute of the quotient space $\Omega/\Gamma$, which
often is a very unpleasant space (e.g., in the case when $\Gamma$
acts ergodically). The general definition of a groupoid is as
follows \cite{Renault-80}.

\begin{definition}
A triple $(\calG,\cdot, ^{-1})$ consisting of a set $\calG$, a
partially defined associative multiplication $\cdot$, and an
inverse operation $^{-1}: \calG \to \calG$ is called a {\em
groupoid} if the following conditions are satisfied:
\begin{itemize}
\item $(g^{-1})^{-1}=g$ for all $g \in \calG$,
\item If $g_1 \cdot g_2$ and $g_2 \cdot g_3$ exist, then $g_1 \cdot g_2
\cdot g_3$ exists as well, \item $g^{-1} \cdot g$ exists always and
$g^{-1} \cdot g \cdot h =  h$, whenever $g \cdot h$ exists, \item $h \cdot
h^{-1}$ exists always and $g \cdot h \cdot h^{-1} = g$, whenever $g \cdot
h$ exists. \end{itemize} \end{definition}

A given groupoid $\calG$ comes along with the following standard
objects. The subset $\calG^0 = \{ g \cdot g^{-1} \mid g \in \calG
\}$ is called the {\em set of units}. For $g \in \calG$ we define
its {\em range} $r(g)$ by $r(g) = g \cdot g^{-1}$ and its {\em
source} by $s(g) = g^{-1} \cdot g$. Moreover, we set $\calG^\omega
= r^{-1}(\{ \omega \})$ for any unit $\omega \in \calG^0$. One
easily checks that $g \cdot h$ exists if and only if $r(h) =
s(g)$.

The groupoids under consideration will always be {\em measurable},
i.e., they posses a $\sigma$-algebra $\calB$ such that all
relevant maps are measurable. More precisely, we require that
$\cdot: \calG^{(2)} \to \calG$, $^{-1}: \calG \to \calG$, $s,r:
\calG \to \calG^0$ are measurable, where
\begin{equation*} \calG^{(2)} := \{ (g_1,g_2) \mid s(g_1) = r(g_2) \} \subset \calG^2 \end{equation*}
and $\calG^0 \subset \calG$ are equipped with the induced
$\sigma$-algebras.  Analogously, $\calG^\omega \subset \calG$ are
measurable spaces with the induced $\sigma$-algebras.

\medskip

As mentioned above, the groupoid associated with (RSM) is simply
$\calG = \Omega \times \Gamma$ and the corresponding operations
are defined as
\begin{equation}        
\label{e-invertieren}
(\omega,\gamma)^{-1} = (\gamma^{-1}\omega,\gamma^{-1}),
\end{equation}
\begin{equation}        
\label{e-multiplizieren}
(\omega_1,\gamma_1)\cdot(\omega_2,\gamma_2) = (\omega_1,\gamma_1
 \gamma_2),
\end{equation}
where the left hand side of \eqref{e-multiplizieren} is only defined if $\omega_1 = \gamma_1 \omega_2$.
It is very useful to consider the elements $(\omega,\gamma)$ of
this groupoid as the set of arrows $\gamma^{-1}\omega
\stackrel{\gamma}{\longmapsto} \omega$. This yields a nice
visualization of the operation $\cdot$ as concatenation of arrows
and of the operation $^{-1}$ as reversing the arrow. The units
$\calG^0 = \{ (\omega,\epsilon) \mid \omega \in \Omega \}$ can 
canonically be identified with the elements of the probability space
$\Omega$. 
Via this identification, the maps $s$ and $r$ assign to
each arrow its origin and its destination.
Our groupoid can be seen as a
bundle over the base space $\Omega$ of units with the fibers
$\calG^\omega = \{ (\omega,\gamma) \mid \gamma \in \Gamma\} \cong
\Gamma$. For simplicity, {\em we
henceforth refer to the set of units as $\Omega$ also in the
setting of an abstract groupoid}. 
The notions associated with the groupoid are illustrated in
Figure 1 for both the abstract case and the concrete case of (RSM).

Next, we introduce an appropriate abstract object which corresponds
to the Riemannian manifold $X$ in (RSM).

\begin{definition} Let $\calG$ be a measurable groupoid with the
previously introduced notations. A triple $(\calX, \pi, J)$ is called
a (measurable) {\em $\calG$-space} if the following properties are
satisfied: $\calX$ is a measurable space with associated
$\sigma$-algebra $\calB_\calX$. The map $\pi: \calX \to \Omega$ is
measurable. Moreover, with $\calX^\omega = \pi^{-1}(\{ \omega \})$,
the map $J$ assigns, to every $g \in \calG$, an isomorphism $J(g):
\calX^{s(g)} \to \calX^{r(g)}$ of measurable spaces with the
properties $J(g^{-1})=J(g)^{-1}$ and $J(g_1 \cdot g_2) = J(g_1) \circ
J(g_2)$ if $s(g_1) = r(g_2)$. 
\end{definition}
Note that a picture similar to Figure 1 exists for a $\calG$-space
$\calX$.

An easy observation is that every groupoid $\calG$ itself is a
$\calG$-space with $\pi = r$ and $J(g)h = g \cdot h$.

The $\calG$-space in (RSM) is given by $\calX = \Omega \times X$
together with the maps $\pi(\omega,x) = \omega$ and
\begin{equation*}  \label{page7}
J(\omega,\gamma): \calX^{\gamma^{-1}\omega} \to \calX^\omega, \quad
J(\omega,\gamma)(\gamma^{-1}\omega,x) = (\omega,\gamma x). 
\end{equation*}

Similarly to the  groupoid $\calG$,  an arbitrary $\calG$-space can be
viewed as a bundle over the base $\Omega$ with fibers $\calX^\omega$.

Our next aim is to exhibit natural measures on
these objects. We first introduce families of measures on the
fibers $\calG^\omega$. In the case of (RSM), this
 can be viewed as an appropriate
generalization of the Haar measure on $\Gamma$.

\begin{definition}
Let $\calG$ be a measurable groupoid and the notation be given as
above.

(a) A {\em kernel} of $\calG$ is a map $\nu: \Omega \to
\calM(\calG)$ with the following properties:
\begin{itemize}
\item the map $\omega \mapsto \nu^\omega(f)$ is measurable for every
$f \in \calF^+(\calG)$,
\item $\nu^\omega$ is supported on $\calG^\omega$, i.e.,
$\nu^\omega(\calG - \calG^\omega) = 0$.
\end{itemize}

(b) A {\em transverse function} $\nu$ of $\calG$ is a kernel
satisfying the following invariance condition
\begin{equation*} \int_{\calG^{s(g)}} f(g \cdot h) d\nu^{s(g)}(h) = \int_{\calG^{r(g)}}
f(k) d\nu^{r(g)}(k) \end{equation*} for all $g \in \calG$ and $f \in
\calF^+(\calG^{r(g)})$.
\end{definition}

In (RSM) the discreteness of $\Gamma$ implies that any kernel $\nu$
can be identified with a function $L \in \calF^+(\Omega\times \Gamma)$
via $\nu^\omega= \sum_{\gamma \in \Gamma}L(\omega,\gamma)
\delta_{(\omega,\gamma)}$. For an arbitrary unimodular group
$\Gamma$, the Haar measure $m_\Gamma$ induces a transverse function
$\nu$ by $\nu^\omega = m_\Gamma$ for all $\omega \in \Omega$ on the
groupoid $\Omega \times \Gamma$ via the identification $\calG^\omega
\cong \Gamma$.

\medskip

In the next definition we introduce appropriate measures on the
base space $\Omega$ of an abstract groupoid $\calG$.

\begin{definition} Let $\calG$ be a measurable groupoid with a
transverse
function $\nu$. A measure $\mu$ on the base space
$(\Omega,\calB_\Omega)$ of units is called {\em $\nu$-invariant}
(or simply invariant, if there is no ambiguity in the choice of
$\nu$) if
\begin{equation*} \mu \circ \nu = (\mu \circ \nu)^\sim, \end{equation*}
where $(\mu \circ \nu)(f) = \int_\Omega \nu^\omega(f)
d\mu(\omega)$ and $(\mu \circ \nu)^\sim(f) = (\mu \circ
\nu)(\tilde f)$ with $\tilde f(g) = f(g^{-1})$.
\end{definition}

\setlength{\unitlength}{1cm}
 \begin{figure}
 \begin{center}
 \psfrag{h}{$h$}
 \psfrag{k}{$k=g \cdot h$}
 \psfrag{G1}{$\calG^{\omega'}=\calG^{s(g)}$}
 \psfrag{G}{$\calG^\omega=\calG^{r(g)}$}
 \psfrag{r}{$r$}
 \psfrag{o1}{$\omega'=s(g)$}
 \psfrag{o3}{$=r(h)$}
 \psfrag{o}{$\omega=r(g)$}
 \psfrag{o2}{$=r(k)$}
 \psfrag{gg}{$g \cdot$}
 \psfrag{g}{$g$}
 \psfrag{Om}{$\Omega$}
 \psfrag{Omm}{$(\Omega,\mu)$}
 \psfrag{GOG}{$\calG = \Omega \times \Gamma$}
 \psfrag{GGa}{$(G^\omega,\nu^\omega) \equiv (\Gamma,m_\Gamma)$}
 \psfrag{gga}{$g = (\omega,\gamma) \cong \left( \omega'
 \stackrel{\gamma}{\longmapsto} \omega \right)$}
 \includegraphics[height=6cm]{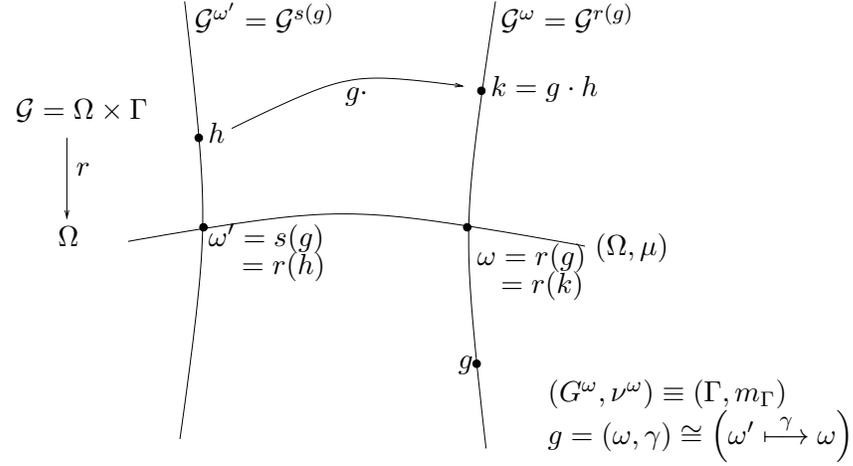}
 \end{center}
 \caption{Notations of the groupoid $\calG = \Omega \times \Gamma$ in
(RSM)}
 \end{figure}

In (RSM) it can easily be checked that, with the above choice $\nu
\equiv m_\Gamma$, a measure $\mu$ on $(\Omega,\calB_\Omega)$ is
$\nu$-invariant if and only if $\mu$ is $\Gamma$-invariant in the
classical sense, see \cite[Cor. II.7]{Connes-1979} as well. Thus a
canonical choice for a $m_\Gamma$-invariant measure on $\Omega$ is
$\PP$. 

Analogously to transverse functions on the groupoid, we introduce a
corresponding fiberwise consistent family $\alpha$ of measures on the
$\calG$-space, see the next definition. We refer to the resulting
object $(\calX,\alpha)$ as a random variable in the sense of Connes.
These random variables are useful substitutes for measurable functions
on the quotient space $\Omega/\Gamma$ with values in $X$. Measurable
functions on $\Omega/\Gamma$ can be identified with $\Gamma$-invariant
measurable functions on $\Omega$. Note that, in the case of an ergodic
action of $\Gamma$ on $\Omega$, there are no nontrivial
$\Gamma$-invariant measurable functions, whereas there are usually
lots of random variables in the sense of Connes (see below for
examples).

\begin{definition}
Let $\calG$ be a measurable groupoid and $\calX$ be a
$\calG$-space. A choice of measures $\alpha: \Omega
\to \calM(\calX)$ is called a {\em random variable} (in the sense
of Connes) with values in $\calX$ if it has the following
properties
\begin{itemize}
\item the map $\omega \mapsto \alpha^\omega(f)$ is measurable
for every $f \in \calF^+(\calX)$,
\item $\alpha^\omega$ is supported on $\calX^\omega$, i.e.,
$\alpha^\omega(\calX - \calX^\omega) = 0$,
\item $\alpha$ satisfies the following invariance condition
\begin{equation*} \int_{\calX^{s(g)}} f(J(g)p) d\alpha^{s(g)}(p) = \int_{\calX^{r(g)}}
f(q) d\alpha^{r(g)}(q) \end{equation*} for all $g \in \calG$ and $f \in
\calF^+(\calX^{r(g)})$. \end{itemize} \end{definition}

To simplify notation, we write $g h$ respectively $g p$ for $g \cdot h$
respectively $J(g) p$.

The general setting for the sequel consists of a groupoid $\calG$
equipped with a fixed transverse function $\nu$ and an
$\nu$-invariant measure $\mu$ on $\Omega$, and a fixed random
variable $(\calX,\alpha)$. We use the follwing notation for 
the ``averaging'' of a $u \in \calF^+(\calX)$ with respcet to $\nu$
\begin{displaymath}
\nu\ast u_0(p) 
:= \int_{\calG^{\pi(p)}} u_0(g^{-1} p) d\nu^{\pi(p)} (g)
\quad \text{ for } p \in \calX. 
\end{displaymath}

We will need the following  further assumptions in order to apply the
integration theory developed in \cite{Connes-1979}.

\begin{definition} \label{admissett}
Let $(\calG,\nu,\mu)$ be a measurable groupoid and $(\calX,\alpha)$ be
a random variable on the associated $\calG$-space $\calX$ satisfying the
following two conditions
\stepcounter{equation}
\begin{enumerate}
\item[\refstepcounter{equation} {\rm (\theequation)} \label{Sprop}]
The $\sigma$-algebras $\calB_{\calX}$ and $\calB_{\Omega}$ are
generated by a countable family of sets, all of which have finite measure, 
w.r.t.~$\mu \circ \alpha$ (respectively w.r.t.~$\mu$). 
\item[\refstepcounter{equation} {\rm (\theequation)} \label{Pprop}]
There exists a strictly positive function $u_0 \in \calF^+(\calX)$ 
satisfying \\$\nu\ast u_0(p) = 1 $ for all $p \in \calX$.
\end{enumerate}
Then we call the tupel $(\calG,\nu,\mu,\calX,\alpha,u_0)$ an {\em
admissible setting}. 
\end{definition}

Before continuing our investigation let us shortly discuss the
relevance of the above conditions: Condition (\ref{Sprop}) is a strong
type of separability condition for the Hilbert space $L^2(\calX,
\mu\circ \alpha)$. It enables us to use the techniques from direct
integral theory discussed in Appendix A which are crucial to the
considerations in Section \ref{vna}.

Condition (\ref{Pprop}) is important to apply Connes' non-commutative
integration theory. Namely, it says that $(\calX,J)$ is proper in the
sense of Lemma III.2 and Definition III.3 of \cite{Connes-1979}.
Therefore $(\calX,J)$ is square integrable by Proposition IV.12 of
\cite{Connes-1979}. This square integrability in turn is a key
condition for the applications of \cite{Connes-1979} we give in
Sections \ref{vna} to \ref{DanielsIDS}.
\smallskip

On an intuitive level, \eqref{Pprop} can be understood as providing an
``embedding'' of $\calG$ into $\calX$. Namely, every $u\in \calF
(\calX)$ with $\nu\ast u\equiv 1$ gives rise to the fibrewise defined
map $q = q_{u}\colon \calF(\calG) \to \calF(\calX)$ by
\begin{equation} \label{qu}
q(f)(p) := \int_{\calG^{\pi(p)}} u(g^{-1}p) f(g) d\nu^{\pi(p)}(g)
\end{equation}
for all $p \in \calX$. Note that the convolution property
of $u$ implies that the map (\ref{qu}) satisfies $q(1_\calG) =
1_\calX$. Moreover, $q$ can be used to obtain new functions $w \in
\calF(\calX)$ satisfying $\nu\ast w \equiv 1$. This is the statement
of the next proposition.

\begin{prop} \label{norm}
 Let $u \in \calF^+(\calX)$ with $\nu \ast u \equiv 1$ be given, and
 $q$ be as above.  For any function $f \in \calF(\calG)$ with
 $\nu(\tilde f) \equiv 1$ on $\Omega$ we have $\nu\ast q(f) \equiv 1$.
\end{prop}

\begin{proof} The proof is given by the following direct
calculation with $\omega = \pi(p) \in \Omega$:
\begin{eqnarray*}
(\nu \ast q(f))(p) &=& \int_{\calG^\omega}
q(f)(g^{-1}p)d\nu^{\omega}(g)\\
&=& \int_{\calG^\omega} \int_{\calG^{s(g)}} u(h^{-1}g^{-1}p)f(h)
 d\nu^{s(g)}(h) d\nu^{\omega}(g)\\
\text{($\nu$ transverse function)} &=& \int_{\calG^\omega}
\int_{\calG^{\omega}} u(k^{-1}p)f(g^{-1}k) d\nu^\omega(k)
d\nu^\omega(g)\\
\text{(Fubini)} &=& \int_{\calG^{\omega}} u(k^{-1}p)
\int_{\calG^{\omega}} f(g^{-1}k) d\nu^\omega(g) d\nu^\omega(k)\\
&=& \int_{\calG^{\omega}} u(k^{-1}p)
\int_{\calG^{\omega}} \tilde f(k^{-1}g) d\nu^\omega(g) d\nu^\omega(k)\\
\text{($\nu(\tilde f)\equiv 1$)} &=& \int_{\calG^{\omega}} u(k^{-1}p)
d\nu^\omega(k)\\ &=& 1. \end{eqnarray*} Note that in the above calculation
the integration variable $g$ has the property $r(g) = \pi(p) = \omega$.
This finishes the proof. 
\end{proof}

\begin{remark} \label{admsett}
We consider examples of admissible settings in the case (RSM). Recall
that we will identify $\calG^\omega$ with $\Gamma$ and
$\calX^\omega$ with $X$ for all $\omega \in \Omega$. A transverse
function of $\calG$ is given by copies of the Haar measure:
$\nu^\omega = m_\Gamma$ for all $\omega \in \Omega$.
$\calX$ together with the Riemannian volume forms $\lambda^\omega$
(corresponding to the metrics $g_\omega$) on the fibers $\calX^\omega
\equiv X$ is an example of a random variable. This will be shown next
by discussing validity of conditions \eqref{Pprop} and \eqref{Sprop}.
\smallskip 

{\em Condition \eqref{Pprop}:} Let $\calD$ be a fundamental domain for
the $\Gamma$ action on $X$ such that $\bigcup_{\gamma \in \Gamma}
\gamma \calD = X$ is a disjoint union, c.f. \cite[\S
6.5]{Ratcliffe-91}. Then every function $v \in \calF(\Gamma)$ with
$\sum_{\gamma \in \Gamma} v(\gamma) = 1$ gives rise to a $u_0$
satisfying $\nu * u_0 \equiv 1$ by
\begin{equation*} u_0(\omega,x) = v(\gamma) \end{equation*} 
where $\gamma \in \Gamma$ is the unique element with $x \in \gamma
\calD$. Note that this construction assigns to a strictly positive
$v$, again a strictly positive $u_0$ on $\calX$. Hence the setting
$(\Omega \times \Gamma,$ $m_\Gamma,\PP,\Omega \times X,\lambda,u_0)$
satisfies condition \eqref{Pprop}.

\smallskip 

{\em Condition \eqref{Sprop}:} This condition is clearly satisfied if
the $\sigma$-algebra of $\Omega$ is countably generated.  In the case
(RSM) (I) this countability condition can always be achieved by
passing to an equivalent version of the defining stochastic
process. Namely, given a random potential $V\colon \Omega \times X \to
\RR$, we construct a stochastic process $\tilde V \colon \tilde \Omega
\times X \to \RR$ with the same finite dimensional distributions such
that \eqref{Sprop} is satisfied,
c.f.~e.g.~\cite{GihmanS-74,PasturF-1992}.

Assume at first that the random potential $V\colon \Omega
\times X \to \RR$ can be written as 
\begin{equation} \label{kirschmodell}
V_\omega(x) = \sum_{\gamma \in \Gamma} f_\gamma (\omega, \gamma^{-1}x).  
\end{equation}
Here $(f_\gamma)_{\gamma \in \Gamma}$ is a sequence of measurable
functions on the probability space $\Omega$ with values in a separable
Banach space $B$ of functions on $X$. Such models have been studied by
Kirsch in \cite{Kirsch-81}. Note that the Borel-$\sigma$-algebra
$\calB_B$ is generated by a countable set $\calE \subset \calB_B$. In
this case $\Omega$ can be chosen to be the canonical probability space
$B^\Gamma$. Its $\sigma$-algebra is generated by a countable family of
cylinder sets of the form
\begin{equation*}
\{\omega \in B^\Gamma | \, \forall \gamma \in H^n : \omega (\gamma)\in
M_\gamma \}
\end{equation*}
where $H$ denotes a finite set of generators of $\Gamma$, $n\in \NN$
and $M_\gamma \in \calE$ for each $\gamma$. An appropriate choice
for the (separable) Banach space $B$ is:
\begin{equation*}
\ell^1(L^p)(X) := \left\{ f \colon X \to \CC \Bigg | \, \int_\Gamma
\left( \int_{\gamma \calD} |f(x)|^p d\lambda(x) \right )^{1/p} 
dm_\Gamma(\gamma) <\infty \right\}
\end{equation*} 
or 
\begin{equation*}
L^p(\ell^1)(X)
:= \left\{ f \colon X \to \CC \Bigg | \, 
 \left( \int_{\calD} \left(\int_\Gamma |f(\gamma^{-1}x)| 
dm_\Gamma(\gamma)\right)^p d\lambda(x) \right )^{1/p} <\infty \right\}.
\end{equation*} 
Here $\calD$ denotes an arbitrary $\Gamma$-fundamental domain in
$X$. We have the inclusion 
\begin{equation*}
L^p(X) \subset \ell^1(L^p(X))\subset L^p(\ell^1(X))
\end{equation*}
and $L^p(\ell^1(X))$ is separable, cf.~\cite{Kirsch-81,Kirsch-89a}. More
correctly, one could use the notation $\ell^1(\Gamma,L^p(\calD))$ for
$\ell^1(L^p)(X)$ and $L^p(\calD,\ell^1(\Gamma))$ for $L^p(\ell^1)(X)$.
\smallskip

Now we show that actually all models which are of the the type (RSM)
(I) can be written in the way (\ref{kirschmodell}). Let $u\colon X \to
\RR^+$ be a bounded measurable function such that $\sum_{\gamma \in \Gamma}
u(\gamma^{-1} x) = 1$ for all $x \in X$. Let $\phi = V u$. Then 
\begin{equation*}
V(\omega,x) = \sum_{\gamma \in \Gamma} V(\omega,x) u(\gamma^{-1}x) = 
\sum_{\gamma \in \Gamma} V(\gamma^{-1}\omega,\gamma^{-1}x) u(\gamma^{-1}x) =
\sum_{\gamma \in \Gamma} \phi(\gamma^{-1}\omega,\gamma^{-1}x).
\end{equation*}   
Setting $f_\gamma(\omega,x) = \phi(\gamma^{-1}\omega,x)$ we have a
representation as in \eqref{kirschmodell}. 

The regularity assumptions on $V$ in (RSM) imply that $V(\omega,
\cdot) \in L^p_{loc, unif}(X) $ for all $\omega \in \Omega$ and all $p
\in [1,\infty]$. 
Here $L^p_{loc,unif}(X)=\ell^\infty(L^p)(X)$ denotes the set of locally
$L^p$-integrable functions $f$ such that the $L^p$-norm of $f
\chi_{\gamma \calD}$ is uniformly bounded in $\gamma \in \Gamma$.
The choice $u = \chi_{\calD}$ yields $\phi(\omega, \cdot)\in L^p(X)$.
Thus the functions $f_\gamma (\omega, \cdot)$ are in the Banach space 
$B=L^p(X)$.  
\smallskip

Summarizing the above considerations, we conclude that (RSM) (I)
satisfies both Condition \eqref{Sprop} and \eqref{Pprop} (after a
suitable modification of the underlying probability space). The case (RSM) (II)
can be treted similarly.
\end{remark}

A crucial fact about the integration of random variables is given in
the following lemma, essentially contained in (the proof of) Lemma
III.1 in \cite{Connes-1979}. We include a proof for the convenience of
the reader.

\begin{lemma}\label{ind}
 Let $\calG$ be a groupoid with transverse function $\nu$
 and $\nu$-invariant measure $\mu$. Let furthermore $\calX$ be a
 $\calG$-space.  \\
 (a) For a given transverse function $\phi$ on $\calG$, the integral
 $\int_\Omega \phi^\omega(f)\,d\mu(\omega)$ does not depend
 on $f \in \calF^+ (\calG)$, provided $f$ satisfies 
 $\nu(\tilde f) \equiv 1$. \\
 (b) For a given random variable $\alpha$ with values in $\calX$ the
 integral $\int_\Omega \alpha^\omega(u)\, d\mu(\omega)$ does not depend on
 $u$, provided $u\in \calF^+ (\calX)$ satisfies $\nu\ast u \equiv 1$.
\end{lemma}

\begin{proof} We prove first part (b). So let
$(\calX, \alpha)$ and $u,u'\in \calF(\calX)$ with $\nu\ast u' \equiv
\nu\ast u \equiv 1 $ be given. Inserting $1 \equiv \nu\ast u'$, we
calculate
\begin{eqnarray*}
\int_\Omega \alpha^{\omega} (u)\,d\mu(\omega) &=& \int_\Omega
\int_{\calX^\omega} u(p)\,\cdot 1 \cdot \, d\alpha^\omega (p)\, d\mu(\omega) \\
(\text{Fubini}) &=& \int_\Omega \int_{\calG^\omega}
\int_{\calX^\omega}
u'(g^{-1} p)\, u(p)\, d\alpha^{\omega}(p)\, d\nu^\omega (g)\,
d\mu(\omega) \\
(\mu\;\: \mbox{inv.})\;\: &=& \int_\Omega \int_{\calG^\omega}
\int_{\calX^{s(g)}} u'(g p')\, u(p') \,d\alpha^{s(g)} (p')
\,d\nu^\omega
(g)\, d\mu(\omega) \\
(\alpha\;\: \mbox{inv.})\;\: &=& \int_\Omega \int_{\calG^\omega}
\int_{\calX^{\omega}} u'( p)\, u(g^{-1}p) \,d\alpha^{\omega} (p)\,
d\nu^\omega (g)\, d\mu(\omega).
\end{eqnarray*}
Another application of Fubini and use of $\nu\ast u \equiv 1$ then
gives $\int_\Omega \alpha^{\omega} (u)d\mu(\omega)=\int_\Omega
\alpha^{\omega} (u')d\mu(\omega)$ and the proof of (b) is finished.

Note that every groupoid $\calG$ is a $\calG$-space $\calX$.
Furthermore, in this case the convolution condition $\nu\ast u
\equiv 1$ translates into $\nu(\tilde f) \equiv 1$. This proves
(a). \end{proof}

Finally,  we introduce, for a given $\calG$-space $\calX$, a new
$\calG$-space $\calX \times_\Omega \calX$. It consists of the set
\begin{equation*} \calX \times_\Omega \calX := \{ (p,q) \mid \pi(p) = \pi(q) \}
\subset \calX \times \calX. \end{equation*} equipped with the induced
$\sigma$-algebra. The corresponding maps $\pi$ and $J$ of $\calX
\times_\Omega \calX$ are defined in an obvious way.  If $\calX$ is
actually a random variable with measures $( \alpha^\omega )_\omega$, then
$\calX \times_\Omega \calX$ inherits the structure of a random
variable by setting 
\begin{equation*}(\alpha \times_\Omega \alpha)^\omega := \alpha^\omega \otimes 
\alpha^\omega.\end{equation*}

\section{The von Neumann Algebra $\calN(\calG,\calX)$}\label{vna}

In this section we discuss how a von Neumann Algebra arises naturally
for an admissible setting $(\calG,\nu,\mu,\calX,\alpha,u_0)$,
cf.~\cite{Connes-1979}.  The random operators we are interested in are
affiliated to this von Neumann algebra.  Recall that a selfadjoint
operator is called {\em affiliated} to a von Neumann algebra if its spectral
family is contained in the von Neumann algebra.

\medskip

A $\calG$-space $\calX$ is, by definition, a bundle over
$\Omega=\calG^{(0)}$ via $\pi : \calX\longrightarrow \Omega$. This
bundle structure of a random variable $(\calX,\alpha)$ induces a
bundle structure of $L^2(\calX, \mu\circ \alpha)$: Using the Fubini Theorem,
 we can associate with $f\in L^2(\calX, \mu\circ \alpha)$  a family
$(f_\omega)_{\omega \in \Omega}$ of functions  
\begin{equation}\label{Faserf}
f_\omega \in L^2(\calX^\omega, \alpha^\omega)\;\: \mbox{such that}\;\:
f(x) = f_{\pi(x)}(x)
\end{equation}
for $\mu$-almost all $\omega \in \Omega$. This way of decomposing will be a key tool
in the sequel. On the technical level, this is very conveniently
expressed using direct integral theory (see e.g.~\cite{Dixmier-1981})
and the fact that there is a canonical isomorphism
\begin{equation}\label{iso}
L^2(\calX, \mu\circ \alpha)\simeq \int_\Omega^{\oplus}
L^2(\calX^\omega, \alpha^\omega) \, d\mu (\omega).
\end{equation} 
As direct integral theory tends to be rather technical, we try to avoid
it in the sequel and rather give direct arguments which, however, are
inspired by the general theory (cf. Appendix A).

A special role will be played by those operators which respect the
bundle structure of $ L^2(\calX, \mu\circ \alpha)$ given by
\eqref{Faserf}. Namely, we say that the (not necessarily bounded)
operator $A: L^2(\calX, \mu\circ \alpha)\longrightarrow L^2(\calX,
\mu\circ \alpha)$ is {\it decomposable} if there exist operators
$A_\omega : L^2(\calX^\omega, \alpha^\omega)\longrightarrow
L^2(\calX^\omega, \alpha^\omega)$ such that $(A f) (x) = (A_\omega
f_\omega) (x)$, for almost every $\omega \in \Omega$. We then write
$A=\int_\Omega^\oplus A_\omega \,d\mu(\omega)$.

\medskip

Let the groupoid $(\calG,\mu,\nu)$ and the random variable
$(\calX,\alpha)$ be as in the last section. For $g\in \calG$, let
the unitary operator $U_g$ be given by
\begin{equation*}
U_g \colon L^2(\calX^{s(g)},\alpha^{s(g)}) \longrightarrow
L^2(\calX^{r(g)},\alpha^{r(g)}),\; U_g f (p) := f(g^{-1}p).
\end{equation*}

A family $(A_\omega)_{\omega\in \Omega}$ of bounded operators
$A_\omega: L^2(\calX^\omega,\alpha^\omega)\to
L^2(\calX^\omega,\alpha^\omega)$ is called a {\em bounded random
operator} if it satisfies :
\begin{itemize}
\item $\omega\mapsto \langle g_\omega, A_\omega f_\omega\rangle$ is
 measurable for arbitrary $f,g\in L^2(\calX, \mu\circ \alpha)$.
\item There exists a $C\geq 0$ with $\|A_\omega\|\leq C$ for almost
all $\omega \in \Omega$.
\item The equivariance condition $A_{r(g)} = U_g A_{s(g)} U_g^*$ for all 
$g\in \calG$ is satisfied.
\end{itemize}
Two bounded random operators $(A_\omega), (B_\omega)$ are called
equivalent, $(A_\omega)\sim (B_\omega)$ if $A_\omega=B_\omega$ for
$\mu$-almost every $\omega\in \Omega$.  Each equivalence class of
bounded random operators $(A_\omega)$ gives rise to a bounded operator
$A$ on $L^2(\calX, \mu\circ \alpha)$ by $A f (p) := A_{\pi(p)}
f_{\pi(p)}$ (cf.~Appendix A). This allows us to identify the class of
$(A_\omega)$ with the bounded operator $A$.  The following is the main
theorem on the structure of the space of random operators.

\begin{theorem} \label{Neum}
The set $\calN(\calG,\calX)$ of classes of 
 bounded random operators is a von Neumann algebra.
\end{theorem}

\begin{proof}  This follows immediately from
\cite[Thm.~V.2]{Connes-1979}. \end{proof}

To get some insight we give a proof of Theorem \ref{Neum} for the
particular case of (RSM). In this example we can canonically
identify $L^2(\calX^\omega,\alpha^{\omega})$ with
$L^2(X,\lambda^\omega)$ and $L^2(\calX,\mu\circ\alpha)$ with
$L^2(\Omega\times X, \PP\circ\lambda)$.  Under this identification
a bounded random operator is just a family of uniformly bounded
operators
\begin{equation*} 
A_\omega: L^2(X,\lambda^\omega) \to L^2(X,\lambda^\omega),\; 
\: \omega\in \Omega 
\end{equation*} 
with $(\omega,x)\mapsto (A_\omega f_\omega) (x)$ measurable and
$\omega\mapsto A_\omega f_\omega \in L^2(X,\lambda^\omega)$ for
every $f\in L^2(\Omega \times X,\PP \circ \lambda)$, satisfying the
equivariance condition
\begin{equation} 
\label{stern} 
A_\omega = U_{(\omega,\gamma)} A_{\gamma^{-1} \omega} U_{(\omega,\gamma)}^*,
\end{equation} 
where we write $U_g$ as $U_{(\omega,\gamma)}$ due to the
product structure of $\calG = \Omega \times \Gamma$.

In this case we can also associate with $(A_\omega)$ an operator $A$ on
the Hilbert space $ L^2(\Omega \times X, \PP \circ \lambda)$ by
setting $ (A f) (\omega, x) \equiv (A_\omega f_\omega) (x)$.  Now,
\eqref{stern} implies
\begin{equation}\label{sternzwei}
A \widetilde{U}_\gamma = \widetilde{U}_\gamma A,
\end{equation}
where $\widetilde{U}_\gamma$ is the unitary operator on $ L^2(\Omega
\times X, \PP \circ \lambda)$ defined by
$$(\widetilde{U}_\gamma f)_\omega(x) = f_{\gamma^{-1}\omega} (\gamma^{-1}x).$$
Conversely, one can show for a decomposable operator $A$ that
\eqref{sternzwei} implies \eqref{stern} $\PP$-almost everywhere. Now, a
suitable averaging procedure and a change of the fibers $A_\omega$
on a set of $\PP$-measure zero yields (\ref{stern}) for all
$\omega$ (cf.~\cite{Lenz-1999} or \cite[p. 88]{Connes-1979}).

It is well known that a bounded operator $A$ on $L^2(\Omega\times
X)$ is decomposable, i.e. 
$A=  \int_\Omega^\bigoplus  A_\omega \ d\PP(\omega)$  
for a
suitable family $(A_\omega)$, if and only if $A$ commutes with the
multiplication operators $M_{h\circ \pi}$ for every $h\in
L^{\infty}(\Omega,\PP)$ (see, e.g., \cite[Thm.~1 in
II.2.5]{Dixmier-1981}). Summarizing these considerations, we conclude
\begin{equation*}
\calN(\Omega \times \Gamma,\Omega \times X)=
\{\widetilde{U}_\gamma \mid \gamma\in  \Gamma\}' \cap\{M_{h\circ
\pi} \mid h\in L^{\infty}(\Omega,\PP) \}',
\end{equation*}
where $S'$ denotes the commutant of a set $S$ of operators.
Obviously, $\{\widetilde{U}_\gamma \mid \gamma\in \Gamma\}$ and $
\{M_{h\circ \pi} \mid h\in L^{\infty}(\Omega,\PP) \}$ are closed under
taking adjoints. Thus, their commutants are von Neumann algebras.
This reasoning shows directly, in the case of (RSM), that the space of
random operators is a von Neumann algebra.

\section{The canonical trace on $\calN(\calG,\calX)$}

In this section we start with an admissible setting
$(\calG,\nu,\mu,\calX,\alpha,u_0)$ and its associated von Neumann
algebra $\calN(\calG,\calX)$ of bounded operators on
$L^2(\calX,\mu \circ \alpha)$. Let $\calN^+(\calG,\calX)$ denote
the set of non negative  selfadjoint operators in $\calN(\calG,\calX)$.
We will show that every operator $A \in \calN^+(\calG,\calX)$
gives rise to a {\em new random variable} $(\calX,\beta_A)$.
Integrating this random variable, we obtain a weight on
$\calN(\calG,\calX)$.  Under certain (mild) assumptions this
weight can be shown to be a trace.

\medskip

We start by associating a transversal function as well as a random
variable with each element in $\calN^+(\calG,\calX)$. In the following,
$q_\omega(f)$ denotes the restriction of $q(f)$ as defined in \eqref{qu} 
to the fiber $\calX^\omega$.

\begin{lemma} \label{trarand}
Let $A\in \calN^+(\calG,\calX)$.\\
(a) Then $\phi_A$, given by $\phi_A^\omega(f) := \tr (A_\omega
M_{q_\omega (f)})$, $f\in \calF( \calG^\omega)$, defines a
transverse function.\\
(b) Then $\beta_A$, given by $\beta_A^\omega(f) := \tr (A_\omega
M_f)$, $f\in \calF( \calX^\omega)$, is a random variable.
\end{lemma}

\begin{proof} This follows by direct calculation using the
equivariance properties of the family $A_\omega$ and the
$q_\omega$. \end{proof}

Let us recall the following definitions. A {\em weight} on a von
Neumann algebra $\calN$ is a map $\tau: \calN^+ \to [0, \infty ]$
satisfying $\tau(A + B)= \tau (A) + \tau(B)$ and $\tau(\lambda
A)=\lambda \tau(A)$ for arbitrary $A,B\in \calN^+$ and $\lambda
\ge 0$. The weight is called {\em normal} if $\tau(A_n)$ converges
to $\tau(A)$ whenever $A_n$ is an increasing sequence of operators (i.e.
$A_n \leq A_{n+1}, n\in \NN$)  converging strongly to $A$. It is called
{\em faithful} if $\tau(A)=0$ implies $A=0$. It is called {\em semifinite}
if $\tau(A)=\sup\{\tau(B): B\leq A, \tau(B)<\infty\}$.  If a weight $\tau$
satisfies $\tau(C C^*)=\tau(C^* C)$ for arbitrary $C\in \calN$ (or
equivalently $\tau(U A U^*)=\tau (A) $ for arbitrary $A\in \calN^+$ and
unitary $U\in \calN$, cf.~ \cite[Cor. 1 in I.6.1]{Dixmier-1981}), it is
called a {\em trace}.

\begin{theorem} \label{snw}
 For $A\in \calN^+(\calG,\calX)$, the expression
\begin{equation*} \tau(A) :=   \int_\Omega \tr (A_\omega^{\frac{1}{2}} M_{u_\omega}
A_\omega^{\frac{1}{2}} ) d\mu(\omega) = \int_\Omega \tr
(M_{u_\omega}^{\frac{1}{2}} A_\omega M_{u_\omega}^{\frac{1}{2}} )
d\mu(\omega)\end{equation*} 
does not depend on $u \in \calF^+(\calX)$ provided $\nu \ast u\equiv 1$.
\begin{enumerate}[(a)]
\item
The map $\tau:\calN^+(\calG,\calX) \longrightarrow [0,\infty]$
is a faithful, semifinite normal weight on $\calN(\calG,\calX)$.
\item
If the groupoid $\calG$ satisfies the {\em freeness condition} 
\begin{equation} \label{Fprop}
r^{-1}(\omega)\cap s^{-1}(\omega)=\{ \omega \} \ \text{for $\mu$-almost
all $\omega\in \Omega$,} 
\end{equation}
then $\tau$ is a trace.
\end{enumerate}
\end{theorem}

\begin{proof}
That $\tau$ is independent of $u$ follows easily from Lemma \ref{ind}
and Lemma \ref{trarand}.\\ (a) The proof that $\tau$ is a normal
weight is straightforward. The proof of the semifiniteness of $\tau$
is not simple and we refer the reader to Theorem VI.2 in
\cite{Connes-1979}. To show that $\tau$ is faithful, assume
$\tau(A)=0$ for $A\in \calN^+(\calG,\calX)$. This implies $\tr
(M_{u_\omega} A_\omega)=0$ for almost every $\omega$. As the trace
$\tr$ is faithful, this implies $M_{u_\omega} A_\omega=0$ for almost
every $\omega$. Choosing a strictly positive function $u$ (e.g. $u =
u_0$), we infer $A_\omega=0$ for almost every $\omega$ and we see that
$\tau$ is faithful.\\ (b) The freeness condition \eqref{Fprop} easily
shows that Corollary VI.7 of \cite{Connes-1979} is applicable and the
statement follows.
\end{proof}

\begin{remark} (i) Property \eqref{Fprop} can be slightly relaxed (cf.
\cite{Connes-1979}). However, some condition of this form is necessary
to prove the trace property in the generality addressed in the
theorem.  It will turn out to be dispensable in many concrete
cases. Indeed, we will see below that $\tau$ is a trace in the case of
(RSM) without any condition of this form. However, when considering   factorial type
properties this condition again plays an important role.\\
(ii) The motivation to call property \eqref{Fprop} a freeness condition is
easily understood in the context of (RSM). Namely, in this case 
\eqref{Fprop} is equivalent to
\begin{equation}
\PP(\omega \mid \gamma^{-1} \omega = \omega) = 0 \ \text{ for all 
$\gamma \in \Gamma \backslash \{ \epsilon \}$}.
\end{equation}
(iii) Given condition \eqref{Fprop} and some ergodicity assumptions (see
the next section), it follows from \cite{Connes-1979} that the von
Neumann algebra is actually a factor. In this case the trace $\tau$ is
unique (up to a scalar factor).
\end{remark}

We are now heading towards an alternative direct proof of part (b) of
Theorem \ref{snw} in the particular case (RSM). This proof will be based on
a study of $\tau$ for certain operators with a kernel. It will in fact
be more general, in that it does not use the freeness condition
\eqref{Fprop} on the action of $\Gamma$ on $\Omega$. Along our way, we
will also find an instructive formula for $\tau$ on these operators.
Let us emphasize that the approach to prove that $\tau$ is a trace
given below is in no way restricted to (RSM)
(cf. \cite{LenzS-2001,LenzS-2002} for related material).

\medskip

An operator $K$ on $L^2(\calX,\mu \circ \alpha)$ is called a {\em
Carleman operator}  (cf.~\cite{Weidmann-1980} for further details) if
there exists a $k \in \calF(\calX \times_\Omega \calX)$ with 
\begin{equation*} k(p,\cdot)
\in L^2(\calX^{\pi(p)},\alpha^{\pi(p)}) \;\:\mbox{for all $p \in \calX$} \end{equation*}
 such that
\begin{equation*} Kf(p) = \int_{\calX^{\pi(p)}} k(p,q) f(q)
d\alpha^{\pi(p)}(q)=:K_{\pi(p)}f_{\pi(p)}(p). \end{equation*}
 This $k$ is called the {\em
kernel} of $K$: Obviously, $K = \int_\Omega^\oplus K_\omega d\mu$.  Let
$\calK$ be the set of all Carleman operators satisfying  for all $g\in \calG$ 
\begin{equation}
\label{carl} k(gp,gq) = k(p,q) \quad \text{for 
$\alpha^{\pi(g)}\times \alpha^{\pi(g)}$ almost all  $p,q$ }. \end{equation}

\begin{prop}
$\calK$ is a right ideal in $\calN(\calG,\calX)$.
\end{prop}

\begin{proof} Obviously,  $K \in \calK$ is
decomposable. Thus, to show that $\calK$ is a subset of
$\calN(\calG,\calX)$, it suffices to show the equivariance
condition. But this is immediate from (\ref{carl}). Thus, it
remains to show that $\calK$ is a right ideal. This follows from a
standard calculation in the theory of Carleman operators:
$$ KAf(p)= K_{\pi(p)}A_{\pi(p)} f_{\pi(p)}(p) = \langle
\overline{k(p,\cdot)}, A_{\pi(p)}f_{\pi(p)}\rangle_{\pi(p)},
$$
where $\langle f_1, f_2 \rangle_\omega = \int_{\calX^\omega}
\overline{f_1(q)} f_2(q) d\alpha^\omega(q)$. This shows that $KA$
has a kernel $k_{KA}$ given by
\begin{equation*} \overline{k_{KA}(p,q)} = \left( A_{\pi(p)}^*
\overline{k(p,\cdot)}\right)(q), \end{equation*} which finishes the proof.
\end{proof}

For a Carleman operator $K$ the expressions $\tau(K K^*)$ and
$\tau(K^* K)$ can directly be calculated.

\begin{prop}  \label{tausch}
Let $K \in \calK$ be given. Then we have
\begin{equation*} \tau(K^* K) = \int_\Omega \int_{\calX^\omega} \int_{\calX^\omega}
u(q) \vert k(p,q) \vert^2 d\alpha^\omega(p) d\alpha^\omega(q)
d\mu(\omega) = \tau(K K^*), \end{equation*} for any $u \in \calF^+(\calX)$
satisfying $\nu \ast u \equiv 1$.
\end{prop}

\begin{proof} We use the well known formula $\tr(A^*A) = \int_{S \times
S} \vert a(x,y) \vert^2 d\lambda(x) d\lambda(y)$ valid for any bounded
operator $A$ with kernel $a$. Note that this formula holds for both $
a\in L^2(S \times S,\lambda \otimes \lambda)$ and $a\notin L^2(S
\times S,\lambda \otimes \lambda)$. In the latter case both sides of
the formula are just infinity. Now, we can calculate as follows:
\begin{eqnarray*}
\tau(K^* K) &=& \int_\Omega \tr(M_{u_\omega}^{1/2} K_\omega^*
K_\omega
M_{u_\omega}^{1/2}) d\mu(\omega)\\
&=& \int_\Omega \int_{\calX^\omega} \int_{\calX^\omega} u(q) \vert
k(p,q)\vert^2 d\alpha^\omega(p) d\alpha^\omega(q) d\mu(\omega).
\end{eqnarray*}
This proves the first equation. To show the second equation, we 
insert $\nu \ast u \equiv 1$ in the first equation, finding
$$\tau (K^\ast K)= \int_{\Omega} \int_{\calX^\omega}
\int_{\calX^\omega} \left( \int_{\calG^\omega}
u(g^{-1}p)d\nu^{\omega}(g) \right) u(q) \vert k(p,q)\vert^2
d\alpha^\omega(q) d\alpha^\omega(p) d\mu(\omega).$$ 
By Fubini and the fact that $K$ is invariant, we infer
$$\cdots = \int_{\calG} \int_{\calX^{r(g)}} \int_{\calX^{r(g)}}
u(g^{-1}p) u(q) \vert k(g^{-1}p,g^{-1}q) \vert^2 d\alpha^{r(g)}(p)
d\alpha^{r(g)}(q) d(\mu\circ\nu)(g).$$ 
Invoking invariance of $\mu$, i.e. $(\mu \circ \nu)^\sim = \mu \circ \nu$, 
we finally obtain
\begin{eqnarray*}
\cdots & = & \int_{\calG} \int_{\calX^{s(g)}} \int_{\calX^{s(g)}}
u(gp) u(q) \vert k(gp,gq) \vert^2 d\alpha^{s(g)}(p) d\alpha^{s(g)}(q)
d(\mu\circ\nu)(g)\\ &=& \int_{\calG} \int_{\calX^{r(g)}}
\int_{\calX^{r(g)}} u(p') u(g^{-1}q') \vert k(p',q') \vert^2
d\alpha^{r(g)}(p') d\alpha^{r(g)}(q') d(\mu\circ\nu)(g).
\end{eqnarray*}
Now, the desired equality follows by reversing the steps from the
preceding calculation. This finishes the proof. \end{proof}

The proposition shows that $\tau$ has the trace property $\tau(A
A^\ast)=\tau (A^\ast A)$  for all Carleman operators $A$. To show that
$\tau$ is actually a trace, we need two more steps. In the first step, we
show that $\tau$ must be a trace if it satisfies this trace property for
sufficiently many operators.  In the second step, we show that the set of
Carleman operators is large.

\begin{lemma}  \label{spur}
Let $\calI$ be a right ideal in a von Neumann algebra $\calN$
acting on a separable Hilbert space $\calH$ with
$\overline{\calI^\ast \calI(\calH)} = \calH$. Let $\tau$ be a
normal weight on $\calN$ satisfying $\tau(A^\ast A)=\tau(A
A^\ast)$ for all $A\in \calI$. Then $\tau$ is a trace.
\end{lemma}

\begin{proof}  It suffices to show $\tau(U A U^\ast)= \tau(A)$ for
arbitrary $A\in \calN_+$ and unitary $U\in \calN$. Let $C(\calI)$
be the norm closure of $\calI^\ast \calI$. Obviously, $\calI^\ast
\calI$ is a dense ideal in $C(\calI)$. By \cite[Prop.
1.7.2]{Dixmier-1977}, there exists then an increasing approximate
identity $(I_\lambda)$ for $C(\calI)$ in $\calI$. As $\calH$ is
separable, we can choose $(I_\lambda)$ to be a sequence $(I_n)$.
By $\overline{\calI^\ast \calI(\calH)} = \calH$, we infer that
$I_n$ converges monotonously to the identity ${\rm Id}$ on
$\calH$. By polarization, every $I_n\in \calI^\ast \calI$ can be
written as
\begin{equation}\label{blab}
I_n=\sum D_{i,n}^\ast D_{i,n} - \sum C_{j,n}^\ast C_{j,n},
\end{equation}
for suitable, finitely many $D_{i,n}, C_{j,n}\in \calI$. Now, let
an arbitrary $D\in \calI$ be given. Using that $\calI$ is a right ideal,
we can calculate \begin{equation}\label{blub} \tau(U A^{\frac{1}{2}}
D^\ast D A^{\frac{1}{2}} U^\ast)= \tau( D A^{\frac{1}{2}} U^\ast U
A^{\frac{1}{2}} D^\ast)= \tau(D A^{\frac{1}{2}} A^{\frac{1}{2}} D^\ast)=
\tau( A^{\frac{1}{2}} D^\ast D A^{\frac{1}{2}}). \end{equation}

Combining \eqref{blub},\eqref{blab} and the fact that $\tau$ is
normal, one can conclude the proof of the Lemma as follows:
$\tau(U A U^\ast)=\lim_{n\to\infty} \tau(U A^{\frac{1}{2}} I_n
A^{\frac{1}{2}} U^\ast)= \lim_{n\to\infty} \tau( A^{\frac{1}{2}}
I_n A^{\frac{1}{2}}) =\tau(A)$. \end{proof}

\begin{prop}  \label{EsTe} Consider (RSM) and define, for $t\geq 0$, the 
operator $S(t)\colon L^2( \Omega \times X) \to L^2( \Omega \times X)$, by
$S(t) f_\omega (x):= (e^{-t \Delta_\omega} f_\omega)
(x)$. Then, $S(t)$ is a selfadjoint bounded random   operator for each
$t\geq 0$. For $t>0$, the operator $S(t)$ belongs to $\calK$.  The
family $t\mapsto S(t)$ is a strongly continuous semigroup. \end{prop}

\begin{proof} Property \eqref{wm} gives immediately the necessary 
measurability of $S(t)$. Moreover, $e^{-t \Delta_\omega}$ is a
selfadjoint contraction for every $\omega \in \Omega$. Therefore,
Fubini easily shows $\|S(t) f\|\leq \|f\|$ for every $f\in L^2( \Omega
\times X)$. Thus, $S(t)$ is a decomposable operator. As its fibres are
selfadjoint, it is selfadjoint as well
(cf.~\cite[Thm.~XIII.85]{ReedS-78} and \cite[Appendix
A.78]{Dixmier-1977}).  The transformation formula $\Delta_\omega =
U_{(\omega,\gamma)} \Delta_{\gamma^{-1} \omega} U_{(\omega,\gamma)}^*$
shows that $S(t)$ satisfies the equivariance property. Thus, $S(t)$ is
indeed a random operator. Using that $t\mapsto e^{-t \Delta_\omega}$
is a strongly continuous semigroup of operators with norm not
exceeding 1 for every $\omega \in \Omega$, we can directly calculate
that $S(t)$ is a strongly continuous semigroup, as well.

For $t>0$,  every $S(t)$ has a kernel $k^t_\omega$ (see, e.g.,
\cite{Chavel-1984} or
\cite{SchY-1994}).  Using selfadjointness and the semigroup property, we
can calculate

\begin{equation*} \int_X \vert k^t_\omega(x,y) \vert^2
d\lambda^\omega(y) = \int_X  k^t_\omega(x,y) k^t_\omega(y,x)
d\lambda^\omega(y) = k^{2t}_\omega(x,x) < \infty. \end{equation*}

This shows that $S(t)$ belongs to $\calK$.  \end{proof}

Now, the following theorem is not hard to prove.

\begin{theorem} In the example (RSM), the map $\tau$ is a normal trace.
\end{theorem} \begin{proof} By Proposition \ref{tausch} and Lemma
\ref{spur}, it suffices to show that $\calK^* \calK L^2(X\times
\Omega)$ is dense in $ L^2( \Omega \times X)$. But, this follows from
the foregoing proposition, which shows that $S( t) f$ converges to $f$
for every $f \in L^2( \Omega \times X)$, where $S(t)\in \calK^*\calK$ by
the semigroup property. \end{proof}

\section{Fundamental results for random operators\label{DanielsIDS}}

In this section we present a comprehensive treatment of the basic
features (P1) -- (P4) of random operators mentioned in the
introduction. This unifies and extends the corresponding known
results about random Schr\"odinger operators in Euclidean space and
random operators induced by tilings. As in the previous section, we
always assume that $(\calG,\nu,\mu,\calX,\alpha,u_0)$ is an admissible
setting.

\medskip

A function $f$ on $\Omega$ is called {\em invariant} if $f\circ r
= f\circ s$. The groupoid $\calG$ is said to be {\em ergodic }
(with respect to $\mu$) if every invariant measurable function $f$ is
$\mu$-almost everywhere constant. This translates, in the
particular case (RSM), to an ergodic action of $\Gamma$ on
$\Omega$.
We will be mostly concerned with decomposable selfadjoint operators on
$L^2(\calX,\mu \circ \alpha)$. 

For a selfadjoint operator $H$, we denote by $\sigma_{disc} (H),
\sigma_{ess} (H), \sigma_{ac}(H), \sigma_{sc}(H)$ and $
\sigma_{pp}(H)$ the discrete, essential, absolutely continuous,
singular continuous, and pure point part of its spectrum respectively.

\begin{theorem}\label{constant}
Let $\calG$ be an ergodic groupoid. Let $H=\int_\Omega^\oplus H_\omega
d\mu (\omega)$ be a selfadjoint operator affiliated to
$\calN(\calG,\calX)$. 
There exist  $\Omega' \subset \Omega$ of full measure and 
 $\Sigma, \Sigma_\bullet\subset \RR$, $ \bullet=disc,ess,ac, sc,pp$, such that
$$ \sigma(H_\omega)=\Sigma, \quad \sigma_\bullet (H_\omega)=
\Sigma_\bullet\quad \mbox{for all $\omega\in \Omega'$} $$ for $\bullet
= disc,ess,ac, sc,pp$.  Moreover, $\sigma(H)=\Sigma$. 
\end{theorem}

Note that $\sigma_{pp} $ denotes the {\em closure} of the set of eigenvalues.
 
\begin{proof} The proof is essentially a variant of well known
  arguments (cf.~e.g.~
  \cite{KunzS-79,KirschM-1982c,Carmona-1986,CyconFSK-1987,CarmonaL-1990}).
  However, as our stetting is different and, technically speaking,
  involves direct integrals with non constant fibres, we sketch a
  proof.

\medskip

Let $\calJ$ be the family of finite unions of open intervals in $\RR$,
all of whose endpoints are rational. Let $\calJ^k$ consist of those
elements of $\calJ$ which are unions of exactly $k$ intervals. Denote
the spectral family of a selfadjoint operator $H$ by $E_H$. It is not
hard to see that $\omega\mapsto \tr E_{H_\omega} (B)$ is an invariant
measurable function for every $B\in \calJ$ (and, in fact, for every
Borel measurable $B\subset \RR$).  Thus, by ergodicity, this map is
almost surely constant. Denote this almost sure value by $f_B$. As
$\calJ$ is countable, we find $\Omega'\subset \Omega$ of full measure,
such that for every $\omega\in \Omega'$, we have $ \tr E_{H_\omega}
(B)=f_B$ for every $B\in \calJ $.  By
\begin{equation}\label{spectrum}
\sigma(H)=\{\lambda\in \RR : E_H (B)\neq 0, \;\:
\mbox{for all $B\in \calJ $  with $\lambda\in B$  }     \},
\end{equation}
and as $\tr$ is faithful, we infer constancy of $\sigma(H_\omega)$ on
$\Omega'$. By a completely analogous argument, using $\sigma_{ess}
(H)=\{\lambda\in \RR : \tr E_H (B)=\infty \; \:\mbox{for all $B\in
\calI$ with $\lambda\in B$}\}$, we infer almost sure constancy of
$\sigma_{ess} (H_\omega)$ and thus also of $\sigma_{disc} (H_\omega)$.

To show constancy of the remaining spectral parts, it suffices to show
measurability of
\begin{equation}\label{meas}
\omega\mapsto \langle g_\omega, E_{H_\omega}^{pp} (B)
g_\omega\rangle_\omega, \;\: \mbox{and} \;\: \omega\mapsto \langle
g_\omega, E_{H_\omega}^{sing} (B) g_\omega\rangle_\omega
\end{equation}
for every $g\in L^2 (X, \mu\circ \alpha)$ and every $B\in
\calJ$. Here, of course, $E_H^{pp}$ and $E_H^{sing}$ denote the
restrictions of the spectral family to the pure point and singular
part of the underlying Hilbert space, respectively.  To show these
measurabilities, recall that for an arbitrary measure $\mu$ on $\RR$
with pure point part $\mu^{pp}$ and singular part $\mu^{sing}$ we have
\begin{equation*}
\mu^{sing} (B)=\lim_{n\to \infty} \sup_{J\in \calJ, |J|\leq n^{-1}}
\mu (B\cap J), \:\;\: \mu^{pp} (B)=\lim_{k\to \infty} \lim_{n\to
\infty} \sup_{J\in \calJ^{k} , |J|\leq n^{-1}} \mu (B\cap J).
\end{equation*}
Here, the first equation was proven by Carmona (see 
\cite{Carmona-1986,CarmonaL-1990, CyconFSK-1987}), and the second follows by a
similar argument. As this latter reasoning does not seem to be in the
literature, we include a discussion in Appendix B. Given these
equalities, \eqref{meas} is an immediate consequence of measurability
of $\omega\mapsto \langle g_\omega, E_{H_\omega}(B)
g_\omega\rangle_\omega$, (which holds by assumption on $H$).  (Note
that instead of considering $\mu^{pp}$ as above, one could have
considered the continuous part $\mu^{c}$ of $\mu$ by a method given in
\cite{CarmonaL-1990}.)

It remains to show the last statement.  Obviously, $E_H (B)=0$ if and
only if $E_{H_\omega} (B)=0$ for almost every $\omega \in
\Omega$. Using this, \eqref{spectrum}, and almost sure constancy of
$\tr E_{H_\omega} (B)$, infer that $\lambda\in \sigma(H)$ if and only
if $E_{H_\omega} (B)\neq 0$ for every $B\in \calJ$ with $\lambda \in
B$ and almost every $\omega\in\Omega$. This proves the last
statement. Alternatively, one could follow the proof of
\cite[Thm.XIII.85]{ReedS-78} which is valid in the case of
non-constant fibres, too.\end{proof}

Recall that a measure $\phi$ on $\RR$ is a {\em spectral measure} for a
selfadjoint operator $H$ with spectral family $E_H$ if, for Borel
measurable $B\subset\RR$, $\phi(B)=0 \Leftrightarrow E_H (B)=0$.

\begin{prop} \label{dandos}
 For a Borel measurable $B$ in $\RR$ and a selfadjoint $H$ affiliated
 to $ \calN(\calG,\calX)$, let $\rho_H (B)$ be defined by $\rho_H (B)
 := \tau(E_H (B))$. Then $\rho_H$ is a spectral measure for
 $H$. Moreover, for a bounded measurable function $F : \RR
 \longrightarrow [0,\infty)$, the equality $\tau(F(H))= \rho_H (F):=
 \int F(x) d\rho_H (x)$ holds.
\end{prop}

\begin{proof}  As $\tau$ is a normal weight, $\rho_H$ is a measure.  As
$\tau$ is faithful, $\rho_H$ is a spectral measure. The last statement
is then immediate for linear combinations of characteristic functions
with non negative coefficients. For arbitrary functions it then
follows after taking suitable monotone limits and using normality of
$\tau$.  \end{proof}

\begin{definition} \label{dofsta}
The measure $\rho_H$ is called {\em (abstract) density of states}.
\end{definition}

\begin{coro} The topological support $\supp(\rho_H) = \{\lambda \in \RR \mid
\rho_H(]\lambda-\epsilon,\lambda+\epsilon[) > 0\ \text{for all}\ \epsilon > 0
\}$ coincides with the spectrum $\sigma(H)$ of $H$ for every selfadjoint $H$
affiliated to $\calN(\calG,\calX)$. If $\calG$ is, furthermore,
ergodic this gives $\supp (\rho_H)=\sigma(H_\omega)$ for almost every
$\omega\in \Omega$.

\end{coro}
\begin{proof} The first statement is immediate as $\rho_H$ is a
spectral measure. The second follows from Theorem
\ref{constant}. \end{proof}

\begin{remark}
In general, $\rho_H$ is {\em not} the spectral measure of the {\em
fibre $H_\omega$}. Actually, there are several examples where the set
\begin{equation*}
\Omega' := \{\omega \in \Omega | \, \rho_H \mbox{ is a spectral
measure for the operator } H_\omega \}
\end{equation*}
has measure zero. We shortly discuss two classes of them (see
\cite{AvronS-1983} as well for related material).

Firstly we consider the case where $(H_\omega)_\omega$ exhibits
localization in an energy interval $I$. This means that for a set
$\Omega_{loc}\subset \Omega$ of full measure
\begin{equation*}
\sigma(H_\omega) \cap I \not= \emptyset \ \mbox{ and } \
\sigma_c(H_\omega) \cap I = \emptyset
\end{equation*}
for all $\omega \in \Omega_{loc}$.  Examples of such ergodic, random
operators on $L^2(\RR^d)$ and $\ell^2(\ZZ^d)$ can be found e.g.~in the textbooks
\cite{CarmonaL-1990, PasturF-1992, Stollmann-2001}. They particularly
include random Schr\"odinger operators in one- and higher dimensional
configuration space.  If $H_\omega$ is an ergodic family of
Schr\"odinger operators on $\ell^2(\ZZ^d)$ or $L^2(\RR)$ one knows moreover for all
energy values $E\in \RR$
\begin{equation*}
\PP \{ \Omega(E) \}=0, \mbox{ where } \Omega(E) := \{ \omega \in
\Omega | \ E \mbox{ is an eigenvalue of } H_\omega \}.
\end{equation*}
Assume that $\rho_H$ is a spectral measure of $H_\omega$ for some $
\omega \in \Omega_{loc}$. Then there exists an eigenvalue $E \in I$ of
$H_\omega$ and consequently $\rho_H(\{E\}) > 0$. Thus $\rho_H$ can only
be a spectral measure for $H_{\omega'}$ if $\omega' \in \Omega (E)$,
but this set has measure zero.
 
Similarly, there are one-dimensional discrete random Schr\"odinger
operators which have purely  singular continuous spectrum almost surely.
An example is the almost Mathieu operator, cf.~\cite{AvronS-1983}, for
a certain range of parameters. Moreover, in \cite{GordonJLS-1997} it
is proven that the singular continuous components of the spectral
measures of these models are almost surely pairwise orthogonal. Thus
again, $\rho_H$ can be a spectral measure only for a set of $\omega$
of measure zero.
\end{remark}

\begin{lemma}\label{key}
Let $\calG$ be ergodic with respect to $\mu$  with  $\mu(\Omega)<
\infty$ such that the following  {\em exhaustion
property} holds:
\begin{align} \label{Eprop}
&\text{There exists a sequence } (f_n) \text{ in } \calF^+(\calG)
\text{ with }
\calG=\bigcup\nolimits_n \{g: f_n(g)>0\},\\
&\|f_n\|_{\infty}\rightarrow 0 \text{ as } n\rightarrow \infty
\text{ and } \nu(\widetilde{f_n}) \equiv 1 \text{ for all } n \in \NN.
\nonumber
\end{align}
Then, for every transverse function $\phi$, either $\phi^{\omega} (1)
\equiv 0$ almost surely  or $\phi^{\omega} (1) \equiv \infty$ almost
surely.
\end{lemma}

\begin{proof}  By ergodicity, $\phi^{\omega} (1)$ is constant
almost surely. Assume $\phi^{\omega} (1)=c<\infty$ for almost
every $\omega$. Lemma \ref{ind} (a) implies that $\int_\Omega
\phi^{\omega} (f)
d\mu(\omega)$ is independent of $f \in \calF^+$ with
$\nu(\widetilde{f})\equiv 1$. For such $f$,  we infer
\begin{equation*} Constant=\int_\Omega \phi^{\omega} (f) d\mu\leq \|f\|_{\infty}\, c\,
\mu(\Omega).\end{equation*} 
As this is in particular valid for every $f_n$, $n\in \NN$, we infer
$Constant =0$.  This shows $\phi^\omega (f_n)=0$ for almost every
$\omega$ and every $n\in \NN$. By $\calG=\bigcup_n \{g: f_n(g)>0\}$,
this gives $\phi^\omega (1)=0$ for almost every $\omega \in \Omega$.
\end{proof}

\begin{remark}
Let $(\calG,\nu,\mu)$ be a measurable groupoid, $(\calX,\alpha)$ be an
associated random variable and $u \in \calF^+(\calX)$ such that property
(\ref{Sprop}) holds and $\nu * u \equiv 1$. Let $f_n$ be a sequence
satisfying condition \eqref{Eprop}. Then $(\calG,\nu,\mu,\calX,\alpha,u_0)$ 
with $u_0 = q_{u}(\sum_{n=1}^\infty \frac{1}{2^n} f_n)$ is an admissible
setting.
\end{remark}
\smallskip 

\begin{remark} The exhaustion property \eqref{Eprop} can easily be
  seen to hold in the case (RSM), if $\Gamma$ is infinite. Namely, we
  can choose
\begin{equation*} f_n (g) = f_n(\omega,\gamma) =\frac{1}{|I_n|} \chi_{I_n}(\gamma). \end{equation*}
Here, $I_n$ is an exhaustion of the group $\Gamma$.
\end{remark}

The lemma has an interesting spectral consequence.

\begin{coro}\label{discrete}
Let the assumptions of Lemma \ref{key} be satisfied and  the
selfadjoint  $H= \int_\Omega^\oplus H_\omega d\mu(\omega)$ be affiliated
to $\calN(\calG,\calX)$.  Then $H_\omega$ has almost surely no discrete
spectrum. \end{coro}

\begin{proof} We already know that the discrete spectrum is constant
almost surely.  Let $B$ be an arbitrary Borel measurable subset of
$\RR$. Then, the map $\omega\mapsto \tr(E_{H_\omega} (B)
M_{q_\omega(\cdot)})$ is a transverse function, by Lemma \ref{trarand}
(a).  Thus, $\tr (E_{H_\omega} (B) M_{q_\omega(1)}) = \tr(
E_{H_\omega}(B) )$ equals almost surely zero or infinity.  As $B$ is
arbitrary, this easily yields the statement.  \end{proof}

\begin{remark} Corollary \ref{discrete} is well known for certain
classes of random operators. However, our proof of the key ingredient,
Lemma \ref{key}, is new. It seems to be more general and more
conceptual. Moreover, as discussed in the proof of Theorem
\ref{typeII} below, Lemma \ref{key} essentially implies that
$\calN(\calG,\calX)$ is not type $I$. Thus, the above considerations
establish a connection between the absence of discrete spectrum and
the type of the von Neumann algebra $\calN(\calG,\calX)$.
\end{remark}

Let us finish this section by discussing factorial and type
properties of $\calN(\calG,\calX)$.  By  \cite[Cor. V.8]{Connes-1979} (cf.
 Cor.  V.7 of \cite{Connes-1979}, as well),  the von Neumann algebra $\calN$
is a factor (i.e. satisfies $\calN \cap \calN'= \CC\, {\rm Id}$) if $\calG$ is
ergodic with respect to $\mu$ and the freeness condition \eqref{Fprop} holds.

\smallskip

There are three different types of factors. These types can be
introduced in various ways. We will focus on an approach centered
around traces (cf.~\cite{Connes-1994} for further discussion and
references).

A factor is said to be of type $III$, if it does not admit a   semifinite
normal trace. If a factor admits such a  trace, then this trace must be
unique (up to a multiplicative constant) and there are two cases. Namely,
either, this trace assumes only a discrete set of values on the projections,
or the range of the trace on the projections is an interval of the form
$[0,a]$ with $0< a\leq\infty$.  In the first case, the factor is said to
be of type $I$. It must then be isomorphic to the von Neumann algebra of
bounded operators on a Hilbert space. In the second case, the factor is
said to be of type $II$.

\begin{theorem} \label{typeII}
Let the assumptions of Lemma \ref{key} and condition (\ref{Fprop}) be
satisfied, then $\calN(\calG,\calX)$ is a factor of  type $II$.
\end{theorem} 
\begin{proof} By Lemma \ref{key}, there does not exist a bounded
transversal function $\phi$ whose support $\{\omega:
\phi^\omega(1)\neq 0\}$ has positive $\mu$ measure. By \cite[Cor.
V.9]{Connes-1979}, we infer that $\calN(\calG,\calX)$ is not type
$I$. On the other hand, as $\tau$ is a semifinite normal trace on
$\calN$ by Theorem \ref{snw}, it is not type $III$.\end{proof}

\begin{remark}
(a) In the case (RSM), under the countability and freeness assumptions
\eqref{Sprop} and \eqref{Fprop}, we know that $\calN(\Omega \times
\Gamma, \Omega \times X)$ is actually a factor of type $II_\infty$:
Since the identity on an infinite dimensional Hilbert space has trace
equal to infinity, we conclude
\begin{equation*} \tr_{L^2(X)}(M_{\chi_\calD}) = \tr_{L^2(\calD)} ({\rm Id}) = \infty, \end{equation*}
where $\calD$ is a fundamental domain as in Remark \ref{admsett}.
Now, choosing $u_\omega (x) \equiv \chi_\calD(x)$, we have $\nu \ast u \equiv 
1$ and, consequently,
\begin{equation*} 
\tau({\rm Id}) = \EE\{ \tr(M_{\chi_\calD}) \} = \infty. 
\end{equation*}  
(b) In the case of tiling groupoids and percolation models one finds factors of type $II$
with a finite value of the canonical trace on the identity. This finite value is  
determined by geometric features of the underlying tiling, respectively the
percolation process, cf.~Sections \ref{Quasicrystal} and  \ref{Percolation}.
\end{remark}

\section{The Pastur-\v Subin-trace formula for (RSM)}
\label{PVresults} 

The aim of this section is to give an explicit exhaustion construction
for the abstract density of states (cf.~Section \ref{DanielsIDS}) of
the model (RSM). The construction shows in particular that the
IDS, the distribution function of the density of states, is
\emph{self-averaging}. This means that it can be expressed by a
macroscopic limit which is $\omega$-independent, although one did note
take the expectation over the randomness.

We recall in this section  the relevant definitions and
results of \cite{LenzPV-03a} concerning the example (RSM)
and outline the main steps of the proofs.

In the following we assume that the group
$\Gamma$ is {\em amenable} to be able to apply an appropriate ergodic theorem.
The exhaustion procedure yields a limiting distribution function which
coincides, at all continuity points, with the distribution function
of the abstract density of states. For the calculation of this distribution
function the Laplace transformation has proved useful
\cite{Pastur-1971,Shubin-1982}. We refer to the second reference for a
detailed description of the general strategy.

In Section 3 of \cite{LenzPV-03a} the operators $H_\omega$ of (RSM) are defined
via quadratic forms, the  measurability of the latter is established, and the  
validity of the measurability condition \eqref{wm} is deduced under the 
following additional hypotheses
\stepcounter{equation}
\begin{enumerate}
\item[{\rm (\theequation)} \label{M1}]
The map $\Omega \times TX \to \RR$, $(\omega,v) \mapsto g_\omega(v,v)$
is jointly measurable.
\item[\refstepcounter{equation} {\rm (\theequation)} \label{M2}]
There is a $C_g \in \, ]0,\infty[$ such that
\begin{equation*}
\label{quasiisom}
C_g^{-1} \, g_0(v,v) \le g_\omega(v,v) \le C_g  \, g_0(v,v) \  \text{ for all } \,
\, v \in TX.
\end{equation*}
\item[\refstepcounter{equation} {\rm (\theequation)} \label{M3}]
There is a $C_\rho  \in \, ]0,\infty[$ such that
\begin{equation*}
\vert \nabla_0 \, \rho_\omega(x) \vert_0 \le C_\rho \  \mbox{for all} \,
\, x \in X,
\end{equation*}
where $\nabla_0$ denotes the gradient with respect to $g_0$, $\rho_\omega$ is the unique smooth density of $\lambda^0$ with
respect to $\lambda^\omega$, and $\vert v \vert_0^2 = g_0(v,v)$.
\item[\refstepcounter{equation} {\rm (\theequation)} \label{M4}]
There is a uniform lower bound $K \in \RR$ for the Ricci curvatures of all
Riemannian manifolds $(X,g_\omega)$. Explicitly, $\Ric (g_\omega )\ge  K g_\omega$
for all $\omega \in \Omega$ and on the whole of $X$.
\item[\refstepcounter{equation} {\rm (\theequation)} \label{M-potential}]
Let $V\colon \Omega \times X\to \RR$ be a jointly measurable mapping such that
for all $\omega\in\Omega$ the \emph{potential} $V_\omega:= V(\omega,\cdot) \ge 0$ 
is in $L^1(A)$ for any compact $A \subset X$.
\end{enumerate} 

%

A key technique is an ergodic theorem by Lindenstrauss
\cite{Lindenstrauss-2001}, valid for amenable groups $\Gamma$ acting
ergodically by measure preserving transformations on $\Omega$.  This
theorem relies on suitable sequences $(I_n )_n$, $I_n \subset \Gamma$,
so called \emph{tempered F{\o}lner sequences} introduced by Shulman
\cite{Shulman-1988}. For an appropriate fundamental domain $\calD$
(cf. \S 3 in \cite{AdachiS-1993}), the sequence $( I_n )_n$ induces a
sequence $( A_n )_n$, $A_n \subset X$ by $A_n = {\rm int}(
\overline{\bigcup_{\gamma \in I_n} \gamma \calD} )$. The ergodic theorem in
\cite{Lindenstrauss-2001} together with the equivariance property 
\eqref{compcomp} imply that the following limits hold pointwise almost surely 
and in $L^1(\Omega,\PP)$ sense
\begin{align} \label{oho}
\lim_{n\to \infty}\frac{1}{| I_n|}\tr(\chi_{A_n}e^{-t H_\omega})
 = \EE\{\tr(\chi_\calD e^{-tH_\bullet})\}
\\
\label{vollim}
\lim_{n\to \infty}\frac{1}{|I_n|}\lambda^\omega(A_n) =\EE(\lambda^\bullet(\calD)).
\end{align}
Moreover the  $L^\infty(\Omega,\PP)$ norms of the sequences 
\begin{equation*}
\Big(\frac{\tr(\chi_{A_n}e^{-t H_\omega})}{| I_n|}\Big)_n,  \quad
\Big(\frac{\lambda^\omega(A_n)}{|I_n|}\Big)_n, \quad
\Big(\frac{| I_n|}{\lambda^\omega(A_n)}\Big)_n
\end{equation*}
are uniformly bounded in the variable $n\in \NN$.

Another important ingredient in the proof of (\ref{Laplace}) below is the
following heat kernel lemma (see \cite[Lemma 7.2]{LenzPV-03a}):
Let $( A_n )_n$ be as above. Then we have 
\begin{equation} \label{alaaf}
\lim_{n \to\infty}\sup_{\omega \in \Omega} \frac{1}{\lambda^\omega(A_n)} \left \vert
\tr(\chi_{A_n} e^{-t H_\omega} )- \tr(e^{-t H_\omega^n}) \right \vert = 0.
\end{equation}

\medskip

The sets $A_n$ together with the random family
$( H_\omega )$ of Schr\"odinger operators on $X$ are used to
introduce the normalized eigenvalue counting functions
\begin{equation*}
N_\omega^n(\lambda) = \frac{| \{ i \mid \lambda_i(H_\omega^n) <
  \lambda\}|} {\lambda^\omega(A_n)},
\end{equation*}
where $H_\omega^n$ denotes the restriction of $H_\omega = \Delta +
V_\omega$ to the domain $A_n$ with Dirichlet boundary conditions and
$\lambda_i(H_\omega^n)$ denotes the $i$-th eigenvalue of $H_\omega^n$
counted with multiplicities. The cardinality of a set is denoted by $|\cdot|$.
Note that $\tr (e^{-tH_\omega^n})= \int e^{-t\lambda} dN_\omega^n (\lambda)$.
Using  \eqref{oho}, \eqref{vollim} and \eqref{alaaf}
it is shown in \cite{LenzPV-03a} that there
exists a distribution function $N_H\colon \RR \to [0,\infty[$, i.e., $N_H$
is left continuous and monotone increasing, such that the following 
$\PP$-almost sure pointwise and $L^1$-convergence of
Laplace-transforms holds true: \ for all $t > 0$
\begin{equation} \label{Laplace}
\lim_{n \to \infty} \tilde N_\omega^n(t) = \lim_{n \to \infty}
\int_{-\infty}^\infty e^{-t \lambda} dN_\omega^n(\lambda) =
\int_{-\infty}^\infty e^{-t \lambda} dN_H(\lambda) = \tilde N_H(t),
\end{equation}
and that $\tilde N_H$ can be identified with the explicit expression
\begin{equation} \label{e-quotient}
\tilde N_H(t) 
= \frac{\EE(\tr(\chi_\calD e^{-tH_\bullet}))}
{\EE(\lambda^\bullet(\calD))}
= \frac{\tau( e^{-tH})}
{\EE(\lambda^\bullet(\calD))}. 
\end{equation}
This implies by the Pastur-\v Subin-Lemma
\cite{Pastur-1971,Shubin-1979} the following convergence (for
$\PP$-almost all $\omega \in \Omega$)
\begin{equation} \label{helau}
\lim_{n \to \infty} N_\omega^n(\lambda) = N_H(\lambda)
\end{equation}
at all continuity points of $N_H$. Note that $N_H$ does not depend on $\omega$.
Moreover, $N_H$ does not depend on the
sequence $( A_n )_n$ as long as $( A_n )_n$ is chosen in the above
way. The function $N_H$ is called the {\em integrated density of states}.

Now, our trace formula reads as follows.

\begin{theorem} \label{pvids}
Let the measure $\rho_H$ be the abstract density of states introduced in
Section \ref{DanielsIDS}. Then we have
\begin{equation*}
N_H(\lambda)
= 
\frac{\rho_H \left (]-\infty,\lambda[ \right ) }{\EE(\lambda^\bullet(\calD)) }
\end{equation*}
at all continuity points $\lambda$ of $N_H$.
\end{theorem}

\begin{remark}
The theorem implies in particular
\begin{equation*} 
N_H(\lambda) 
= \frac{1}{\EE(\lambda^\bullet(\calD))} \EE\left\{  \tr
\left( \chi_\calD E_{H_\omega}(]-\infty, \lambda[) \right)\right\} , 
\end{equation*}   
where $\calD$ is a fundamental domain of $\Gamma$ as in Remark \ref{admsett}.
This alternative localized formula for the IDS is well-known in the
Euclidean case. Note that it doesn't rely on a choice of boundary
condition.
\end{remark}

\begin{proof} By the uniqueness lemma for the Laplace transform (see
Lemma \ref{unicus} in the Appendix) it suffices to show that for all $t >0$
\begin{eqnarray}
\label{Laplace=} 
\EE(\lambda^\bullet(\calD)) \int e^{-t \lambda} d N_H(\lambda) 
= \int e^{-t \lambda} d \rho_H (\lambda ) .
\end{eqnarray}

To this end we observe that by \eqref{e-quotient}
$\tau(e^{-t H}) = \EE(\lambda^\bullet(\calD))
\int e^{-t \lambda} d N_H(\lambda)$
which leaves to prove that 
\begin{equation}\label{rest}
\tau(e^{-t H}) = \rho_H (e^{-t \lambda}). 
\end{equation}

To do so, we will use that the operator $H$ is bounded below, say
$H\geq C$, with a suitable $C\in \RR$. Define $F:\RR\longrightarrow
[0,\infty[$ by $F(\lambda)= e^{-t\lambda}$ if $\lambda\geq C$ and by
$F(\lambda)=0$ otherwise. By spectral calculus, we infer $e^{-tH}=
F(H)$ and, in particular, $\tau(e^{-t H})= \tau(F(H))$. By Proposition
\ref{dandos}, this implies $\rho_H (F)=\tau (e^{-t H})$. As, again by
Proposition \ref{dandos}, $\rho_H$ is a spectral measure for $H$, its
support is contained in $[C,\infty[$ and we easily find
$\rho_H(e^{-t\lambda})=\rho_H (F)$. Combining these equalities, we end
up with the desired equality \eqref{rest}.
\end{proof}

\section{Quasicrystal models} 
\label{Quasicrystal}
In this section we shortly discuss how to use the above framework to
study random operators associated to quasicrystals. Quasicrystals are
usually modelled by tilings or Delone sets and these two approaches
are essentially equivalent. Here, we work with Delone sets and follow
\cite{LenzS-2001,LenzS-2002, LenzS-2003} to which we refer for further
details. The investigation of quasicrystals via groupoids goes back to
Kellendonk \cite{Kellendonk-1995, Kellendonk-1997} and his study of
K-theory and gap labelling in this context (see
\cite{BellissardHZ-2000,LenzS-2001, LenzS-2002,LenzS-2003} for further
discussion of quasicrystal groupoids and
\cite{BellissardKL-01,BenameurOH-01,BenedettiBG-2001,KaminkerP-2003}
for recent work proving the so-called gap-labelling conjecture).

\medskip

A subset $\omega$ of $\RR^d$ is called {\em Delone} if there exist $0 < r
\leq R$ such that $r \leq \| x - y\|$ whenever $x,y\in \omega$ with
$x\neq y$ and $\omega \cap\{ y : \|y - x\|\leq R\}\neq \emptyset$ for
all $x\in \RR^d$. Here, $\|\cdot\|$ denotes the Euclidean norm on $\RR^d$. 

There is a natural action $T$ of $\RR^d$ on the set of all Delone sets
 by translation (i.e. $T_t \omega = t + \omega$). Moreover, there is a
 topology (called the natural topology by some authors) such that
 $T$ is continuous. Then, $(\Omega,T)$ is called a Delone dynamical
 system if $\Omega$ is a compact $T$-invariant set of Delone sets.

In this case $\calG(\Omega,T):=\Omega\times \RR^d$ is clearly a
groupoid with transversal function $\nu$ with $\nu^\omega
=\mbox{Lebesgue measure}$ for all $\omega\in \Omega$. If $\mu$ is a
$T$-invariant measure on $\Omega$, it is an invariant measure on
$\calG(\Omega,T)$ in the sense discussed above.  By the compactness of
$\Omega$ there exists at least one such non trivial $\mu$ by the
Krylov-Bogolyubov theorem. In fact, in the prominent examples for
quasicrystals, there is a unique such probability measure; these
systems are called {\em uniquely ergodic}.  This notation comes from
the fact that this unique $T$-invarant measure is necessarily ergodic.

We now assume that $(\Omega,T)$ with invariant measure $\mu$ is given. Then, 
there is a natural space $\calX$  given by
\begin{equation*}\calX=\{(\omega,x)\in \calG(\Omega,T) : x\in \omega\}\subset \calG(\Omega,T).\end{equation*} 

Then,  $\calX$  inherits a topology from $\calG(\Omega,T)$ and is in fact
a closed subset. The space $\calX$ is  fibred over $\Omega$ with
fibre map
\begin{equation*}
\pi :  \calX \longrightarrow \Omega, \;\:(\omega,x)\mapsto \omega.
\end{equation*}
Thus, the fibre $\calX^\omega$ can naturally be identified with
$\omega$. In particular, every $g = (\omega,x)\in \calG(\Omega,T)$ gives
rise to a isomorphism $J(g) : \calX^{s(g)}\to \calX^{r(g)}$,
$J(g)(\omega-x,p)= (\omega,p +x)$ and  $J(g_1 g_2)=J(g_1)
J(g_2)$ and $J(g^{-1})=J(g)^{-1}$. Each fibre $\omega$  carries the discrete
measure $\alpha^\omega$ giving the weight one to each point of $\omega$. Then, $(\calX,\alpha)$ is a random variable.

Let furthermore $u \geq 0$ be a continuous function on $\RR^d$ with
$\int u(t) dt = 1$. Then, $u$ gives rise to a function $u_0$ on $\calX$ via
\begin{equation*}u_0 (\omega,x) = u(x)\end{equation*}
and
\begin{equation*}
  \int_{\calG(\Omega,T)^{\pi (p)}} u_0 (\gamma^{-1} p
  ) d\nu^{\pi (p)} (\gamma) = \int_{\RR^d} u(t) dt =1.
\end{equation*} 
Therefore, we are in an admissible setting.

The freeness condition that $\gamma^{-1} \omega \neq \omega$ whenever
$\gamma \neq (\varepsilon, \omega)$ is known as \textit{aperiodicity}.

The associated operators are given by families $A_\omega : \ell^2
(\omega) \longrightarrow \ell^2 (\omega)$, $\omega \in \Omega$
satisfying a measurability and boundedness assumption as well as the
equivariance condition
\begin{equation*} 
A_{\omega} = U_{(\omega,t)} A_{\omega-t} U_{(\omega,t)}^\ast
\end{equation*} 
for
$\omega \in \Omega$ and $t\in \RR^d$, where $U_{(\omega,t)} : \ell^2
(\omega-t) \longrightarrow \ell^2 (\omega)$ is the unitary operator
induced by translation.

There is a canonical trace on these operators given by
\begin{equation*}
\tau(A) :=\int_\Omega \tr (M_{u_0} A_\omega)d \mu (\omega). 
\end{equation*} 
It is not hard to see that $\tau ({\rm Id})< \infty$ as there exists
$r>0$ with $\|x - y\| \geq r$ whenever $x,y\in \omega$ for $\omega \in
\Omega$ and $x\neq y$. If $\mu$ is ergodic, then $\tau ({\rm Id})$ is
just the density of points of almost all $\omega \in \Omega$.  In this
case, we can conclude from the discussion in Section 5 that the discrete
spectrum is absent. The necessary sequence $f_n$ is defined by
\begin{equation*} 
f_n (\omega,x) = \frac{1}{{\rm vol}(B_n)}\chi_{B_n} (x),
\end{equation*} 
where $B_n$ is
the ball in $\RR^d$ around the origin with radius $n$, $\chi$ denotes
the characteristic function and ${\rm vol}$ stands for Lebesgue measure.

In fact, assuming ergodicity of $\mu$ together with aperiodicity, 
i.e.~freeness, we can even conclude from Section 5 that the von Neumann
algebra of random operators is a factor of type $II_1$.

Of course, in the ergodic case the results of Section \ref{DanielsIDS}
can be applied. They give almost sure constancy of the spectral components
and the possibility to express the spectrum of a random operator $A$
with the help of the measure $\rho_A$ defined there by
$\rho_A(\varphi) = \tau (\varphi (A))$ for continuous $\varphi$ on
$\RR$ with compact support.

\begin{theorem}
\label{QC-Spectralproperties} 
Let $(A_\omega)$ be a selfadjoint random operator in the setting
discussed in this section and asume that $\mu$ is ergodic. 
Then there exists $\Omega' \subset \Omega$ of full measure and
subsets of the real numbers $\Sigma$ and $ \Sigma_\bullet$, where
$\bullet \in\{disc, ess, ac, sc, pp\}$, such that for all $\omega\in
\Omega'$
\begin{equation*}
  \sigma(A_\omega)=\Sigma \quad \text{ and } \quad \sigma_\bullet (A_\omega)= 
\Sigma_\bullet
\end{equation*}
for any $\bullet = disc, ess, ac, sc, pp$. 
Moreover, $\Sigma_{disc}=\emptyset$ and  $\Sigma$ coincides with the
topological support of $\rho_A$. 
\end{theorem}

\medskip

In the situation of the theorem it is also possible to calculate the
distribution function of $\rho_A$ by a limiting procedure.  Details
are discussed in the literature cited above, see \cite{LenzS-2003,
  LenzS-2002}.  Here, we mention the following results for so
\emph{called finite range operators} $(A_\omega)$: Denote by $|\cdot|
$ the number of elements of a set and by $A_\omega|_{B_n}$ the
restriction of $A_\omega$ to $\omega\cap B_n$.  Define the measures
$\mu_\omega^n$ on $\RR$ by
\begin{equation*}
C_0(\RR) \, \ni \, \varphi \mapsto \mu_\omega^n (\varphi) 
:= \frac{1}{|B_n \cap\omega|} \tr \, \varphi (A_\omega|_{B_n}).
\end{equation*} 
These are just the  measures associated to  the eigenvalue
counting functions studied so far, i.e. 
\begin{equation*}
 N_\omega^n(\lambda) := \mu_\omega^n (]-\infty,\lambda[)= 
\frac{|\{i : \lambda_i
  (A_\omega|_{B_n}) < \lambda \}|}{|B_n \cap\omega|}
\end{equation*}
with the (ordered) eigenvalues $\lambda_i (A_\omega|_{B_n})$ of
$A_\omega|_{B_n}$.  Set $D:=\tau ({\rm Id})$. 
Then, the measures $\mu_\omega^n$ converge for $n\to \infty$ vaguely to 
the measure
\begin{equation*}
  \varphi \mapsto \frac{1}{D} \tau ( \varphi (A ) ) = \frac{1}{D}   
  \rho_A (\varphi)
\end{equation*} 
for almost every $\omega\in \Omega$. If $(\Omega,T)$ is uniquely ergodic, the
convergence holds for all $\omega\in \Omega$. Assuming further regularity, one
can even show convergence of the corresponding distribution functions with
respect to the supremum norm. In any case there exists a distribution
function $N_A$ such that $\lim_{n \to \infty} N^n_\omega(\lambda) = 
N_A(\lambda)=\frac{1}{D} \rho_A
(]-\infty,\lambda[)$ exists almost surely at all continuity points of $N_A$.

Note that amenability is not an issue here since the group $\RR^d$ is
abelian. In fact, instead of the balls $B_n$ we could also consider
rather general van Hove sequences.

\bigskip

Let us finish this section by giving an explicit example of what may
be called a nearest neighbour Laplacian in the context of a Delone set
$\omega$.  For $x\in \omega$, define the \emph{Voronoi cell}
$\calV(x)$ of $x$ by
\begin{equation*} \calV(x) := \{p\in\RR^d : \|p - x\| \leq \|p - y\|
  \;\:\mbox{for all $ y\in \omega$ } \}.\end{equation*} Then, it is
not hard to see that $\calV(x)$ is a convex polytope for every $x\in
\omega$. The operator $A_\omega$ is then defined via its matrix
elements by
\begin{equation*} A_\omega (x,y) = 1 \text{ if $\calV(x)$ and
    $\calV(y)$ share a $(d-1)$-dimensional face}\end{equation*} and
$A_\omega (x,y)=0$ otherwise.

\section{Percolation models}\label{Percolation}

In this section we shortly discuss how to fit percolation operators,
more precisely site percolation operators, in our framework. 
In fact, edge percolation or mixed percolation could be treated along the same lines.  

Theoretical physicists have been interested in Laplacians on percolation graphs as quantum mechanical 
Hamiltonians for quite a while \cite{deGennesLM-59a,deGennesLM-59b,KirkpatrickE-72,ChayesCFST-86}. 
Somewhat later several computational physics papers where devoted to 
the numerical analysis of spectral properties of percolation Hamiltonians,
see e.g~\cite{ShapirAH-82,KantelhardtB-97,KantelhardtB-98,KantelhardtB-98b,KantelhardtB-02}.
More recently there was a series of rigorous mathematical results on percolation models
\cite{BiskupK-01a,Veselic-05a,Veselic-05b,KirschM-04,MuellerS}.

\medskip

We have to identify the abstract quantities introduced in the abstract setting in the context of percolation.
\medskip

A graph $G$ may be equivalently defined by its vertex set $X$ and its
edge set $E$, or by its vertex set $X$ and the distance function $d \colon
X\times X \to \{0\} \cup \NN$. We choose the second option and tacitly
identify the graph with its vertex set. In particular, each graph $X$ 
gives naturally rise to the so called
\emph{adjacency matrix} $A(X) : X\times X\longrightarrow \{0,1\}$ defined by
$A(X) (x,y) =1$ if $d(x,y) =1$ and $A(X)(x,y)=0$ otherwise. This
adjacency matrix can be considered to be an operator on $\ell^2 (X)$.

\medskip

We assume that the vertex set of the graph is countable. Let $\Gamma$
be a group acting freely on $X$ such that the quotient $ X / \Gamma$
is a finite graph, i.e. the $\Gamma$-action is quasi-transitive.  The
associated probability space is given by $\Omega=\{0,1\}^X$, with the
$\sigma$-field defined by the finite-dimensional cylinder sets.  For
simplicity let us consider only independent, identically distributed
percolation on the vertices of $X$.  The statements hold analogously
for correlated site or bond percolation processes under appropriate ergodicity
assumptions.  Thus we are given a sequence of i.i.d.~random variables
$\omega_x \colon \Omega \to \{0,1\}$, for $x \in X$, with distribution
measure $p \delta_1 + (1-p) \delta_0$.  The measure $\PP$ on the
probability space $\Omega$ is given by the product $\PP=\bigotimes_ X
(p \delta_1 + (1-p) \delta_0)$.
  
The groupoid is given by $\calG := \Omega \times \Gamma \ni g
=(\omega, \gamma)$ and the $\calG$-space by
\begin{equation*}
\calX:= \Omega \times X.  
\end{equation*} 
The operation of the groupoid is the same as defined in
\eqref{e-invertieren} and \eqref{e-multiplizieren}, the projection
$\pi$ is again given by $\pi(\omega, x):= \omega$ for all $(\omega ,x) \in
\Omega \times X$, and the map $J$ is the same as described on page
\pageref{page7} for the model (RSM), where $x$ now stands for a vertex of the graph $X$.
\bigskip

Define for $\omega\in \Omega$ the random subset 
\begin{equation*}
X(\omega) := \{ x \in X \mid \omega_x =1\} 
\end{equation*}
of $X$. Defining $\alpha$ by $\alpha^\omega:= \delta_{X(\omega)}$ 
for all $\omega \in \Omega$ we obtain a random variable $(\calX,\alpha)$.

\medskip

Similarly as in the case of random Schr\"odinger operators on manifolds
we define the transverse function $\nu$ on the groupoid by setting $\nu^\omega$ equal to the counting measure $m_\Gamma$ 
on the group for every $\omega\in \Omega$ and 
the $\nu$-invariant measure $\mu$ equal to $\PP$.
\bigskip

We next show that we are in an admissible setting according to
Definition \ref{admissett}.  As for the countability condition in this
definition, we proceed as follows: By the countability of the vertex
set of the graph $X$ the $\sigma$-field $\calB_\calX$ is countably
generated. Since $\calB_\Omega$ is generated by the finite dimensional
cylinder sets, it is countably generated as well.
 
Denote by $\calD$ a fundamental domain of the covering graph $X$
and set $u_0 (\omega,x) := \chi_\calD(x)$. Then clearly
\begin{equation*} \int_{\calG^{\pi(p)}} u_0 (g^{-1} p)d\nu^{\pi(p)}(g) = 
\sum_{\gamma \in \Gamma} u_0 (\omega, \gamma x) = 1 
\quad \text{ for all } \omega \in \Omega, x \in X.
\end{equation*}
This shows the second condition of Definition \ref{admissett}. 

\medskip


The corresponding random operators are given by families $H_\omega
\colon \ell^2 (X,\alpha^\omega) \rightarrow \ell^2 (X,\alpha^\omega)$,
$\omega \in \Omega$, satisfying a measurability and a boundedness
assumption as well as the equivariance condition
\begin{equation*}
H_{\omega} = U_{(\omega,\gamma)} H_{\gamma^{-1}\omega} 
U_{(\omega,\gamma)}^\ast
\end{equation*}
for $\omega \in \Omega$ and $\gamma\in \Gamma$, where
$U_{(\omega,\gamma)} \colon \ell^2 (X,\alpha^{\gamma^{-1}\omega})
\rightarrow \ell^2(X, \alpha^{\omega})$ is the unitary operator
mapping $\phi$ to $\phi(\gamma^{-1} \, \cdot)$. A special example of
such a random operator is given by the family $(A_\omega)_\omega$,
where each $A_\omega$ is the adjacency matrix of the induced subgraph
of $X$ generated by the vertex set $X(\omega)$.

\medskip

We next show that the functional 
\begin{equation*}
\tau(H) := \EE\{  \tr (M_{u_0} H_\omega) \}
\end{equation*}  
defined in Theorem \ref{snw} is a trace. As (a) of this Theorem holds by general arguments, 
we just have to show that the freeness condition \eqref{Fprop} 
is satisfied. This can be shown as follows:
Since $\Gamma$ acts freely, $\gamma x \neq x$ for all $x \in X, \gamma \in \Gamma \setminus \{\epsilon\}$.
Thus the set $\{\omega \mid \ \omega_{\gamma x} =\omega_{x}\}$ has measure smaller 
than one for all $x\in X$. If the graph is infinite, this poses infinitely
many independent conditions on the elements in 
\begin{equation*}
\Omega_\gamma := \{\omega \mid \gamma \omega=\omega\}.
\end{equation*}
Therefore, $\Omega_\gamma$ has measure zero for each $\gamma\in \Gamma\setminus  \{\epsilon\}$.  
As $\Gamma$ is countable, 
$ \bigcup_{\gamma\in \Gamma\setminus \{\epsilon\} } \Omega_\gamma $
has $\PP$-measure zero as well. Thus the desired freeness statement follows.

\medskip

To show that an associated operator family $\omega\mapsto H_\omega$
has almost surely no discrete spectrum for infinite $\Gamma$, we
define the sequence of functions $f_n$ by
\begin{equation*}
f_n(\omega,\gamma) :=  \frac{1}{|I_n|}\chi_{I_n} (\gamma).
\end{equation*}
Here, $I_n$ is an exhaustion of the infinite group $\Gamma$.
The  sequence $(f_n)$ satisfies property \eqref{Eprop}. 
Thus, Lemma \ref{key} and Corollary \ref{discrete} hold.

In fact, the above exhaustion and freeness properties allow us to
apply Theorem \ref{typeII} and conclude that the von Neumann algebra
is of type $II$. More precisely
\begin{equation*}
\tau({\rm Id}) = \EE \{ \tr (\chi_\calD)\} = p |\calD|< \infty
\end{equation*}   
shows that the type is $II_1$.
\medskip


\smallskip

Besides the discrete, essential, absolutely continuous, singular continuous, 
and pure point spectrum,  $\sigma_{disc}, \sigma_{ess}, \sigma_{ac},
\sigma_{sc}, \sigma_{pp}$,  the set  $\sigma_{fin}$ consisting  of eigenvalues 
which posses an eigenfunction with finite support 
is a quantity which may be associated with the whole family 
$H_\omega, \omega \in \Omega$.


The following two theorems hold for a class of percolation Hamiltonians 
$(H_\omega)_\omega$ which are obtained from a determinstic
\emph{finite hopping range operator} by a percolation process,
cf.~\cite{Veselic-05a}.
In particular the class contains the adjacency operator $(A_\omega)_\omega$
introduced above.

Recall from Section \ref{DanielsIDS} that the measure $\rho_H$ on $\RR$ is
given by $\rho_H (\varphi) = \tau (\varphi (H))$ for continuous $\varphi$ on
$\RR$ with compact support.

\begin{theorem}
There exists an $\Omega' \subset \Omega$ of full measure and subsets
of the real numbers $\Sigma$ and $ \Sigma_\bullet$, where $\bullet
\in\{disc, ess, ac, sc, pp, fin\}$, such that for all $\omega\in
\Omega'$
\begin{equation*}
  \sigma(H_\omega)=\Sigma \quad \text{ and } \quad \sigma_\bullet 
  (H_\omega)= \Sigma_\bullet
\end{equation*}
for any $\bullet = disc, ess, ac, sc, pp, fin$. Moreover, the
almost-sure spectrum $\Sigma$ coincides with the topological support
of $\rho_H$.  If $\Gamma$ is infinite, $\Sigma_{disc}=\emptyset$.
\end{theorem}

If the group $\Gamma$ acting on $X$ is amenable
it was shown in \cite{Veselic-05a} that 
property (P5), too,  holds for the family $(H_\omega)_\omega$.
 More precisely, it is possible to construct the
measure $\rho_H$ by an exhaustion procedure using the finite volume
eigenvalue counting functions $N_\omega^n(\lambda)$.  These are
defined by the formula
\begin{equation*}
  N_\omega^n(\lambda)
  := \frac{\nr{\{i \in \NN\, \mid  {\lambda}_i(H_\omega^n) < \lambda \}}}
  { \nr{I_n}\cdot \nr{\calD}} .
\end{equation*} 
where $I_n$ is a tempered F{\o}lner sequence,
$A_n = \bigcup_{\gamma \in I_n} \gamma \calD$,
and $H_\omega^n$ is the restriction 
of the operator $H_\omega$ to the space $\ell^2(X(\omega) \cap A_n)$.

\begin{theorem}
\label{t-exIDS}
Let $\Gamma$ be amenable and $\{ I_n \}$ be a tempered  F{\o}lner sequence
of $\Gamma$. Then there exists a subset $\Omega' \subset \Omega$ of full 
measure and a distribution function $N_H$, called \emph{integrated density of 
states}, such that for all $\omega \in \Omega'$
\begin{equation} 
\label{t-d-IDS}
\lim_{n\to \infty} N_\omega^n (\lambda) = N_H(\lambda),
\end{equation}
at all continuity points of $N_H$. $N_H$ is related to the measure $\rho_H$
via the following \emph{trace formula}
\begin{equation} 
\label{e-Gtraceformula} 
N_H(\lambda) = \frac{\rho_H(]-\infty,\lambda[)}{|\calD|}.
\end{equation}
\end{theorem}

\appendix

\section{Some direct integral theory}
The aim of this appendix is to prove the following lemma and to
discuss some of its consequences. By standard direct integral theory
\cite{Dixmier-1981}, the lemma is essentially equivalent to the
statement that $L^2(\calX,\mu\circ \alpha)$ is canonically isomorphic
to $\int_\Omega^\oplus L^2(\calX^\omega, \alpha^\omega) \,d\mu
(\omega)$.

\medskip

Throughout this section let a measurable groupoid $(\calG,\nu,\mu)$ and
a random variable $(\calX,\alpha)$ satisfying condition (\ref{Sprop}) on
the associated $\calG$-space be given.

\begin{lemma}\label{dit} 
There exists a measurable function $N: \Omega\longrightarrow
\NN_0\cup \{\infty\}$, a sequence $( g^{(n)} )_n$ of measurable
functions on $\calX$ and a subset $\Omega'$ of $\Omega$ of full measure
satisfying the following:
\begin{itemize}
\item $( g^{(n)}_\omega : 1\leq n\leq N(\omega) )$ is an orthonormal
basis of $L^2 (\calX^\omega, \alpha^\omega)$ for every $\omega\in
\Omega'$.
\item $g^{(n)}_\omega = 0$ for $n > N(\omega)$, $\omega\in \Omega'$.
\item $g^{(n)}_\omega = 0$ for $\omega\notin \Omega'$. 
\end{itemize}
\end{lemma}
\begin{proof} Let $\calD$ be a countable generator of the
$\sigma$-algebra of $\calX$ such that $\mu\circ \alpha (D)< \infty$
for every $D\in \calD$. Such a $\calD$ exists by condition
\ref{Sprop}. Let $(f^{(n)} )_{n\in \NN}$ be the family of
characteristic functions of sets in $\calD$. By assumption on $\calD$,
we infer that $(f^{(n)})$ are total in $L^2 (\calX, \mu\circ \alpha)$.

By the Fubini Theorem, there exist a set $\Omega'$ of full measure,
such that $f^{(n)}_\omega$ belongs to $L^2
(\calX^\omega,\alpha^\omega)$ for every $n\in \NN$ and every
$\omega\in \Omega'$. As $\calD$ generates the $\sigma$-algebra of
$\calX$ and $\calX^\omega$ is equipped with the the induced
$\sigma$-algebra, we infer that $(f^{(n)}_\omega )_{n\in \NN}$ is
total in $L^2 (\calX,\alpha^\omega)$ for every $\omega\in \Omega'$.

Now, define for $n\in \NN$ the function $h^{(n)}\in L^2 (\calX,
\mu\circ \alpha)$ by setting $h^{(n)} (p)= f^{(n)} (p)$ if $\pi(p)\in
\Omega'$ and $h^{(n)} (p)=0$, otherwise.

Applying the Gram-Schmidt-orthogonalization procedure to
$(h^{(n)}_\omega )_{n\in \NN}$ simultaneously for all $\omega\in
\Omega'$, we find $N : \Omega \longrightarrow \NN_0 \cup \{\infty\}$
and $g^{(n)}_\omega$ as desired. (This simultaneous Gram-Schmidt
procedure is a standard tool in direct integral theory, see
\cite{Dixmier-1981} for details.) The proof of the Lemma is finished.
\end{proof}

\begin{prop}
Let $(A_\omega)$ be a family of bounded operators $A_\omega \colon L^2
(\calX^\omega, \alpha^\omega) \longrightarrow L^2 (\calX^\omega,
\alpha^\omega)$ such that
\begin{equation}
\label{Sternchen}
 \omega \mapsto \langle f_\omega , A_\omega \, g_\omega \rangle_\omega
 \, \mbox{ is measurable for arbitrary $f,g\in L^2 (\calX,\mu\circ
 \alpha)$}.
\end{equation}
Then, $ Ah \colon \calX \longrightarrow \CC$, $(A h) (p) \equiv
(A_{\pi(p)} h_{\pi(p)})(p)$ is measurable for every $h :
\calX\longrightarrow \CC$ measurable with $h_\omega \in L^2
(\calX^\omega, \alpha^\omega)$ for every $\omega \in \Omega$.
\end{prop}

\begin{proof} 
Let $h$ be be as in the assumption. Invoking suitable cutoff
procedures, we can assume without loss of generality
\begin{equation}
\label{L2echt}
h \in L^2(\calX, \mu \circ \alpha) \text{ as well as } \| A_\omega\| \le C 
\text{ for all } \omega \in \Omega,
\end{equation}
for a suitable $C$ independent of $\omega$.  Obviously,
(\ref{Sternchen}) implies that $\omega \mapsto \langle f_\omega,
A_\omega g_\omega\rangle_\omega$ is measurable for every $f,g :\calX
\longrightarrow \CC$ measurable with $f_\omega, g_\omega \in L^2
(\calX^\omega, \alpha^\omega)$ for every $\omega \in \Omega$.  Thus,
with $g^{(n)}$, $n\in \NN$, as in the previous lemma, we see that
\begin{equation*} \omega\mapsto \langle g^{(n)} _\omega, A_\omega h_\omega\rangle
\:\;\mbox{is measurable for every $n\in \NN$}.\end{equation*} 
In particular,
\begin{equation*} \calX \longrightarrow \CC, \:\; p \mapsto \langle g^{(n)}_{\pi(p)},
A_{\pi(p)} h_{\pi(p)}\rangle_{\pi(p)} \ g^{(n)} (p) \:\;\mbox{is
measurable for every $n\in \NN$}.
\end{equation*} 
As $( g^{(n)}_\omega : 1\leq n \leq N(\omega) )$ is an orthonormal
basis in $L^2 (\calX^\omega, \alpha^\omega)$ for every $\omega\in
\Omega'$ and $g^{(n)}_\omega = 0$ for $n > N(\omega)$ and $\omega\in
\Omega'$, we have
\begin{equation*}
 (A_{\omega} h_\omega) (p) = \sum_{n=1}^{\infty} \langle
 g^{(n)}_\omega, (A_\omega h_\omega)\rangle_\omega \,g^{(n)} (p)
\end{equation*}
for almost every $\omega\in\Omega$. Note that the last equality holds
in the $L^2(\calX^\omega, \alpha^\omega)$-sense.  By (\ref{L2echt})
and the Fubini Theorem, this implies $ (A_{\omega} h_\omega) (p) =
\sum_{n=1}^{\infty} \langle g^{(n)}_\omega, (A_\omega
h_\omega)\rangle_\omega \,g^{(n)} (p)$ in the sense of $L^2(\calX, \mu
\circ \alpha)$ and the desired measurability follows.
\end{proof}

\begin{prop}  Let $C>0$ and $(A_\omega)$ be a family of operators 
$A_\omega\colon L^2 (\calX^\omega, \alpha^\omega) \longrightarrow L^2
(\calX^\omega, \alpha^\omega)$ with $\|A_\omega\|\leq C$ for every
$\omega \in \Omega$ and $p\mapsto (A_{\pi(p)} f_{\pi(p)}) (p)$
measurable for every $f\in L^2 (\calX, \mu \circ \alpha)$. Let $A :
L^2 (\calX, \mu \circ \alpha)\longrightarrow L^2 (\calX, \mu \circ
\alpha)$,\; $(A f) (p) = (A_{\pi(p)} f_{\pi(p)}) (p)$ be the
associated operator. Then, $A_\omega =0 $ for $\mu$-almost every
$\omega\in \Omega$ if $A=0$.
\end{prop} 
{\it Proof.}  Choose $g^{(n)}$, $n\in \NN$, as in Lemma \ref{dit} and
let $\calE$ be a countable dense subset of $L^2 (\Omega,\mu)$.  For $f\in
L^2 (\calX, \mu \circ \alpha)$ with $f(p) = g^{(n)} (p) \psi(\pi(p))$ for
$n\in \NN$ and $\psi \in \calE$, we can then calculate
\begin{equation*} 0 = Af = (p\mapsto \psi(\pi(p)) (A_{\pi(p)} g^{(n)}_{\pi(p)})(p)).\end{equation*}
As $\calE$ is dense and countable, we infer, for $\mu$-almost all
$\omega\in \Omega$,
\begin{equation*} A_\omega g^{(n)}_\omega = 0 \;\:\mbox{for all $n\in \NN$}.\end{equation*} This
proves the statement, as $(g^{(n)}_\omega : n\in \NN)$ is total in
$L^2 (\calX^\omega, \alpha^\omega)$ for almost every $\omega \in
\Omega$. \hfill \qedsymbol

\begin{coro} Let $(A_\omega)$ and $(B_\omega)$ be random operators with 
associated operators $A$ and $B$ respectively. Then $A=B$ implies
$(A_\omega)\sim (B_\omega)$.
\end{coro}
{\it Proof.} This is immediate from the foregoing proposition.\hfill
\qedsymbol

\section{A Proposition from measure theory}

In this appendix we give a way to calculate the point part of a finite
measure on $\RR$.  Recall that a measure is called continuous if it
does not have a point part.

We start with the following Lemma.

\begin{lemma} Let $\mu$ be a continuous finite measure on $\RR$. Then 
$\lim_{n\to \infty} \mu (I_n)=0$ for every sequence $(I_n)$ of open
intervals whose lengths tend to zero.
\end{lemma}
\begin{proof} Assume the contrary. Then there exists a sequence of open
intervals $(I_n)$ with $| I_n | \to 0$ and a $\delta>0$ with
$\mu(I_n)\geq \delta$, $n\in \NN$. For each $n\in \NN$ choose an
arbitrary $x_n\in I_n$.

If the sequence $(x_n)$ were unbounded, one could find a subsequence
$(I_{n_k})_{k\in \NN}$ of $(I_n)$ consisting of pairwise disjoint
intervals. This would imply the contradiction $ \mu(\RR)\geq
\sum_{k=1}^\infty \mu(I_{n_k})\geq \sum_{k=1}^\infty \delta =\infty$.

Thus, the sequence $(x_n)$ is bounded and therefore contains a
converging subsequence. Without loss of generality we assume that
$x_n\to x$ for $n\to \infty$. For every open interval $I$
containing $x$ we then have $\mu(I)\geq \mu(I_n)$ for $n$ large
enough. This gives $\mu(I)\geq \delta$ for every such interval. From
Lebesgue Theorem, we then infer $\mu(\{x\})\geq \delta$, contradicting the
continuity of $\mu$. \end{proof}

\begin{prop} Let $\mu$ be a finite measure on $\RR$ with point part 
$\mu^{pp}$ and continuous part $\mu^{c}$. Then, for every $B\subset
\RR$, $\mu^{pp} (B)$ is given by
\begin{equation*}\mu^{pp} (B)= \lim_{k\to \infty} \lim_{n\to \infty} \sup_{|J|\leq
n^{-1}, J\in \calJ^k} \mu(B\cap J).\end{equation*}
\end{prop} 

\begin{proof} Obviously, the limits on the right hand side of the
formula make sense. We show two inequalities:

``$\leq$'': Let $\{x_i\}$ be a countable subset of $\RR$ with
$\mu^{pp} = \sum_{i} \mu(\{x_i\}) \delta_{x_i}$, where $\delta_x$
denotes the point measure with mass one at $x$. Then, we have
$\mu^{pp} (B)=\sum_{x_i\in B} \mu(\{x_i\})$. For every $\epsilon>0$, we
can then find a finite subset $B_\epsilon$ of $\{x_i : x_i\in B\}$
with number of elements $\# B_\epsilon$ and $\mu^{pp} (B) \leq
\epsilon + \sum_{x\in B_\epsilon} \mu(\{x\})$. This easily gives
\begin{equation*}\mu^{pp} (B) \leq \epsilon + \sum_{x\in B_\epsilon} \mu(\{x\}) \leq
\epsilon + \mu(B \cap I)\end{equation*} 
for suitable $J\in \calJ^{\# B_\epsilon}$ of arbitrary small Lebesgue
measure. As $\epsilon$ is arbitrary, the desired inequality follows.

\medskip

``$\geq$'': By the foregoing lemma, we easily conclude for every $k
\in \NN$ that $\lim_{n\to \infty} \sup_{J\in \calJ^k, |J|\leq n^{-1}}
\mu^{c}(B\cap J) = 0$. Combining this with the obvious inequality
$\mu^{pp} (B) \geq \mu^{pp} (B\cap J)$ valid for arbitrary measurable
$B,J\subset \RR$, we infer
\begin{eqnarray*}
\mu^{pp} (B) &\geq& \lim_{k\to \infty} \lim_{n\to \infty}
\sup_{|J|\leq n^{-1}, J\in \calJ^k} \mu^{pp} (B\cap J)\\ &=
&\lim_{k\to \infty} \lim_{n\to \infty} \sup_{|J|\leq n^{-1}, J\in
\calJ^k} \mu^{pp} (B\cap J) + \lim_{k\to \infty} \lim_{n\to \infty}
\sup_{|J|\leq n^{-1}, J\in \calJ^k} \mu^{c} (B\cap J)\\ &= &\lim_{k\to
\infty} \lim_{n\to \infty} \sup_{|J|\leq n^{-1}, J\in \calJ^k}
\mu(B\cap J).
\end{eqnarray*}
This finishes the proof. \end{proof}

\section{Uniqueness lemma for the Laplace transform}

The results in \cite{PeyerimhoffV-2001} use heavily the Laplace
transform techniques developed in papers by Pastur and \v Subin
\cite{Pastur-1971,Shubin-1982}. In the present paper only the
uniqueness lemma is used.  In the literature the uniqueness lemma
for the Laplace transform is stated mostly for finite measures
(e.g., Theorem 22.2 in \cite{Billingsley-1995}).  For the
convenience of the reader we show how to adapt the uniqueness
result to our case, where the distribution function $N_H$ is
unbounded.

\begin{lemma} \label{unicus}
\label{unique} Let $f_1, f_2  \colon  ]0 , \infty [ \to \RR$ be
monotonously increasing function with $ \lim_{\lambda \searrow 0}
f_1 (\lambda) =\lim_{\lambda \searrow 0} f_2 (\lambda) =0$. Let
the integrals
\begin{equation}
\int_0^\infty e^{-t \lambda} d f_j(\lambda) , \ j = 1,2 , \ t > 0
\end{equation}
be finite and moreover
\begin{equation}
\int_0^\infty e^{-t \lambda} d f_1(\lambda) = \int_0^\infty e^{-t
\lambda} d f_2(\lambda)
\end{equation}
for all positive $t$.

Then the sets of continuity points of $f_1$ and $f_2$ coincide and
for $\lambda_0$ in this set we have $f_1(\lambda_0 ) = f_2 (\lambda_0)$.
\end{lemma}

\begin{proof} Choose $s>0$ arbitrary.  The measures
\begin{equation}
\mu_j (g) :=  \int_0^\infty g(\lambda) \, e^{-s \lambda} \, df_j
(\lambda)
\end{equation}
are finite, since $\mu_j(1) = \int_0^\infty e^{-s \lambda} \, df_j
(\lambda) < \infty$ by assumption. Since
\begin{equation}
\mu_j (e^{-t \cdot}) =  \int_0^\infty \, e^{-(t+s) \lambda} \,
df_j (\lambda) = \tilde f_j (t+s),
\end{equation}
we have, by assumption, that the Laplace transforms of the
measures $\mu_j$ coincide for all $t>0$:
\begin{equation}
\mu_1 (e^{-t \cdot}) = \mu_2 (e^{-t \cdot}).
\end{equation}
As we are dealing with finite measures, the Theorem 22.2 in
\cite{Billingsley-1995} implies $\mu_1 = \mu_2$. We consider
\begin{equation}
\mu_j ([0,E]) = \int_0^E  e^{-s \lambda} \, df_j(\lambda)
\end{equation}
as a sequence of integrals depending on the parameter $s \to 0$.
Since
\begin{equation}
e^{-s \cdot} \colon [0,\lambda_0] \to [0,1]
\end{equation}
converges uniformly and monotonously to the constant function $1$, 
we conclude by Beppo Levi's theorem
\begin{equation}
\lim_{s\searrow 0} \mu_j ([0,\lambda_0]) = \int_0^{\lambda_0} df_j(\lambda).
\end{equation}
For a continuity point $\lambda_0$ of $f_1$ we have
\begin{equation}
\int_0^{\lambda_0} df_1 (\lambda) = f_1(\lambda_0),
\end{equation}
which implies $ f_1(\lambda_0)= f_2(\lambda_0)$. \end{proof}

\begin{coro}
Under the assumptions of the Lemma \ref{unique} we have
\begin{equation}
\int_0^\infty \, g(\lambda) \, df_1 (\lambda) =
 \int_0^\infty \, g(\lambda) \, df_2 (\lambda)
\end{equation}
for all continuous functions $g$ with compact support.
\end{coro}

\end{document}